\pdfoutput=1

\documentclass[11pt,twoside,a4paper,cmspaper,final,collab]{cms-tdr}
\def\svnVersion{477646P}\def\cmsCernNoTag{CERN-EP-2017-218}\def\cmsCernDate{\today}\def\cmsMessage{Published in Physical Review C as \href{http://dx.doi.org/10.1103/PhysRevC.98.044902}{\doi{10.1103/PhysRevC.98.044902.}}}

\begin{document}\cmsNoteHeader{HIN-15-008}

\hyphenation{had-ron-i-za-tion}
\hyphenation{cal-or-i-me-ter}
\hyphenation{de-vices}
\RCS$Revision: 477644 $
\RCS$HeadURL: svn+ssh://svn.cern.ch/reps/tdr2/papers/HIN-15-008/trunk/HIN-15-008.tex $
\RCS$Id: HIN-15-008.tex 477644 2018-10-10 13:43:59Z qwang $
\newlength\cmsFigWidth
\ifthenelse{\boolean{cms@external}}{\setlength\cmsFigWidth{0.85\columnwidth}}{\setlength\cmsFigWidth{0.4\textwidth}}
\ifthenelse{\boolean{cms@external}}{\providecommand{\cmsLeft}{top\xspace}}{\providecommand{\cmsLeft}{left\xspace}}
\ifthenelse{\boolean{cms@external}}{\providecommand{\cmsRight}{bottom\xspace}}{\providecommand{\cmsRight}{right\xspace}}
\ifthenelse{\boolean{cms@external}}{\providecommand{\NA}{\ensuremath{\cdots}}}{\providecommand{\NA}{\ensuremath{\text{---}}}}
\newcommand{\roots}{\ensuremath{\sqrt{s}}}
\newcommand{\deta}{\ensuremath{\Delta\eta}}
\newcommand{\dphi}{\ensuremath{\Delta\phi}}
\newcommand{\pp}{\ensuremath{\Pp\Pp}\xspace}
\newcommand{\Pb}{\ensuremath{\text{Pb}}\xspace}
\newcommand{\PbPb}{\ensuremath{\text{PbPb}}\xspace}
\newcommand{\pPb}{\ensuremath{\Pp\text{Pb}}\xspace}
\newcommand{\dAu}{\ensuremath{\text{dAu}}\xspace}
\newcommand{\pAu}{\ensuremath{\Pp\text{Au}}\xspace}
\newcommand{\noff}{\ensuremath{N_\text{trk}^\text{offline}}\xspace}
\providecommand{\EPOS}{\textsc{epos}\xspace}
\newcommand{ \dmean }[1]{\ensuremath{\langle\langle #1 \rangle\rangle}}
\newcommand{ \cn }[1]{c_n\{#1\}}
\newcommand{ \vtwo }[1]{\ensuremath{v_2\{#1\}}}
\renewcommand{\Re}{\operatorname{Re}}

\cmsNoteHeader{HIN-15-008}
\title{Pseudorapidity and transverse momentum dependence of flow harmonics in \pPb and \PbPb collisions}

\date{\today}

\abstract{
Measurements of azimuthal angular correlations are presented for high-multiplicity pPb collisions
at $\sqrtsNN = 5.02$\TeV
and peripheral PbPb
collisions at $\sqrtsNN = 2.76$\TeV. The data used in this work were collected with
the CMS detector at the CERN LHC.
Fourier coefficients as functions of transverse
momentum and pseudorapidity are studied using  the scalar product method, 4-, 6-, and 8-particle
cumulants, and the Lee--Yang zeros technique.
The influence of event plane
decorrelation is evaluated using the scalar product method and found to
account for most of the observed pseudorapidity dependence.
}

\hypersetup{%
pdfauthor={CMS Collaboration},%
pdftitle={Pseudorapidity and transverse momentum dependence of flow harmonics in pPb and PbPb collisions},%
pdfsubject={CMS},%
pdfkeywords={CMS, physics, heavy ion, flow}}

\maketitle

\section{Introduction}
\label{sec:intro}

High energy density matter with quark and gluon degrees of freedom, a state of matter known as the quark-gluon plasma (QGP), is created
in relativistic heavy ion collisions at the BNL RHIC and at the
CERN LHC~\cite{Arsene:2004fa, Adcox:2004mh, Back:2004je, Adams:2005dq, Aad:2010bu, Chatrchyan:2011sx}.
The energy density created in the initial heavy ion collision is azimuthally nonuniform
as a consequence of the collision geometry and its fluctuations.
Interactions among constituents in the QGP convert this nonuniformity
into an observable anisotropy in the final-state particle momentum distribution.
The azimuthal angle distribution of emitted particles can be
characterized by its Fourier components~\cite{Voloshin:1994mz}. In particular,
the second and third Fourier components, $v_2$ and $v_3$, known as elliptic and triangular flow, respectively,
most directly reflect the medium response to the initial collision geometry and
its fluctuations~\cite{Alver:2010gr}. The magnitudes of these components provide insights into the fundamental transport
properties of the medium~\cite{Alver:2010dn,Schenke:2010rr,Qiu:2011hf}.
Two-particle correlations in
the azimuthal angle ($\phi$) and pseudorapidity ($\eta$) differences between the two particles ($\dphi$ and $\deta$)
have played a vital role in the observation of the azimuthal
anisotropies~\cite{Adams:2005ph, Alver:2009id, Abdelwahab:2014cvd, Abelev:2009jv, Chatrchyan:2011eka, Aamodt:2011by, Chatrchyan:2012wg, ATLAS:2012at}.
These particle correlations are characterized
by a pronounced structure at $\abs{\dphi} \approx 0$ extending over a large $\deta$ range
(referred to as the ``ridge'').  In collisions between two heavy nuclei, such as
CuCu and AuAu collisions at RHIC~\cite{Adams:2005ph, Alver:2009id, Abdelwahab:2014cvd} and
\PbPb collisions at the LHC~\cite{Aamodt:2011by, Chatrchyan:2011eka, ATLAS:2012at, Chatrchyan:2012wg}, these long-range correlations are often attributed to
the collective flow from a strongly interacting, expanding medium~\cite{Ollitrault:1992bk,Reisdorf:1997fx}.
This is corroborated by multiparticle correlations, suggesting a hydrodynamic origin for the observed
azimuthal anisotropies~\cite{Gale:2013da}.

The lightest systems in which ridge-like structures have been observed include high-multiplicity
final states in \pp~\cite{Khachatryan:2010gv,Aad:2015gqa,Khachatryan:2015lva,Khachatryan:2016txc,Aaboud:2016yar} and
\pPb~\cite{CMS:2012qk, Abelev:2012ola, Aad:2012gla, Aad:2014lta, Aaij:2015qcq, Aaboud:2016yar} collisions at the LHC.
Evidence of such long-range correlations is also observed at a nucleon-nucleon center-of-mass energy of $\sqrtsNN = 200\GeV$
in \pAu~\cite{Aidala:2016vgl}, dAu~\cite{Adare:2013piz,Adare:2014keg,Adamczyk:2015xjc}
and $^{3}$HeAu collisions~\cite{Adare:2015ctn} at RHIC.
In \pPb collisions, the overall strength of the correlation is observed so far to be
significantly larger than in \pp collisions, and is comparable to that found in
peripheral \PbPb collisions~\cite{Chatrchyan:2013nka,Khachatryan:2015waa}.

Both the ATLAS~\cite{Aad:2013fja,Aaboud:2017blb} and CMS~\cite{Chatrchyan:2013nka} experiments
have measured significant elliptic flow coefficients in \pPb collisions at
$\sqrtsNN = 5.02\TeV$ using four-particle correlations
based on the cumulant method~\cite{Borghini:2001vi}.
The long-range correlations persist in measurements that study
the correlation among six or more particles in \pPb collisions~\cite{Khachatryan:2015waa,Khachatryan:2016txc,Aaboud:2017acw}
and in measurements of four-particle and six-particle  correlations in \pp collisions at $\roots = 13\TeV$~\cite{Khachatryan:2016txc, Aaboud:2017blb}.
Four-particle correlation measurements in the \dAu system at $\sqrtsNN = 200$, 62.5, 39, and 19.6\GeV
by the PHENIX Collaboration and a six-particle correlation measurement by
the same collaboration at $\sqrtsNN = 200\GeV$ also find significant elliptic flow coefficients~\cite{Aidala:2017ajz}.

In combination, these measurements support a collective origin of the azimuthal correlations, and have raised
the possibility that a QGP droplet might be formed in small-system collisions
exhibiting fluid-like
behavior~\cite{CMS:2012qk,Abelev:2012ola,Aad:2012gla,Bozek:2012gr,Khachatryan:2015waa}.
If such a mechanism can be confirmed, it will significantly extend the range of system size for which the
QGP medium is considered to exist.
However, the origin of the ridge phenomenon
in small collision systems is still being actively investigated. In addition to a
hydrodynamic origin~\cite{Bozek:2011if,Bozek:2012gr}, possible alternative explanations
include gluon saturation in the initial interacting state of the
protons~\cite{Dusling:2012wy,Dusling:2012cg},
multiparton interactions~\cite{Alderweireldt:2012kt},
and the anisotropic escape of partons from the surface of the interaction region~\cite{He:2015hfa}.

To provide further constraints on the theoretical understanding of the azimuthal anisotropies
in different collision systems, this paper presents results on the pseudorapidity and
transverse momentum dependence of the flow harmonics in \pPb and \PbPb collisions. The $v_2$ coefficients are measured
using the 4-, 6-, and 8-particle Q-cumulants~\cite{Bilandzic:2010jr}, the Lee--Yang zeros (LYZ)~\cite{Bhalerao:2003xf},
and the scalar product methods~\cite{Adler:2002pu,Luzum:2012da}. The $v_3$ coefficients, which result from
fluctuations in the collision geometry, are studied with the scalar product method.
Within the hydrodynamic picture, the longer lifetime of the medium
on the Pb-going side in \pPb collisions is expected to lead to
larger values for both the $v_2$ and $v_3$
flow harmonics than on the p-going side~\cite{Bozek:2015swa}.
The \pPb system is studied at $\sqrtsNN = 5.02\TeV$ using data obtained
by the CMS experiment in 2013.
A sample of \PbPb collision data at$\sqrtsNN = 2.76\TeV$ is also analyzed.
The particle correlations are studied for high-multiplicity \pPb collisions
whose particle densities are comparable to those in mid-central
(50--60\% centrality) \PbPb collisions.
The centrality variable is defined as a fraction of the inelastic hadronic cross section in heavy ion
collisions, with 0\% corresponding to the most central, \ie, head-on collisions.
This allows for a direct comparison of \pPb and \PbPb systems over
a broad range of similar particle multiplicities,
thereby helping to clarify the underlying mechanism responsible for the observed correlations.

\section{The CMS experiment}

A detailed description of the CMS detector can be found in Ref.~\cite{Chatrchyan:2008zzk}.
The results in this paper are mainly based on the silicon tracker detector and two hadron forward calorimeters (HF) located on either side of the tracker.
Situated inside the 3.8 T field of a super-conducting solenoid, the silicon tracker consists of 1\,440 silicon pixel and 15\,148 silicon strip detector modules.
It measures charged particles within the range of $\abs{\eta}<2.4$
and provides an impact parameter resolution of $\approx$15\micron and a \pt resolution better than 1.5\% at $\pt \approx 100 \GeVc$.
Electromagnetic (ECAL) and hadron (HCAL) calorimeters are also located inside the solenoid and cover the range of $\abs{\eta} < 3.0$.
The HCAL has sampling calorimeters composed of brass and scintillator plates.
The ECAL consists of lead-tungstate crystals arranged in a quasi-projective geometry.
Iron/quartz-fiber Cherenkov HF cover the range $2.9 < \abs{\eta} < 5.2$ on either side of the interaction region.
The HF calorimeters, which are used in the scalar product analysis,  are azimuthally subdivided into $20^{\circ}$ modular wedges and further segmented to form $0.175\times10^{\circ}$ $(\Delta\eta{\times} \Delta\phi)$ towers.
The CMS detector response is determined through Monte Carlo (MC) studies using {\GEANTfour} \cite{Agostinelli:2002hh}.
\section{Event and track selection }

The \pPb data set corresponds
to an integrated luminosity of 35\nbinv.
The beam energies were 4\TeV for protons and 1.58\TeV
per nucleon for lead nuclei, resulting in
$\sqrtsNN = 5.02\TeV$. The beam directions were reversed during the run. The results from both beam directions are combined using the convention that the proton-going direction defines positive pseudorapdity.
As a result of the energy difference between the colliding beams, the nucleon-nucleon
center-of-mass frame in the \pPb collisions is not at rest with respect to the laboratory frame.
Massless particles emitted at $\eta_\text{c.m.} = 0$ in the nucleon-nucleon center-of-mass frame will be detected at $\eta =  0.465$  in the laboratory frame. Unless otherwise stated, all pseudorapidities reported in this paper are referred to with respect to the laboratory frame.
A sample of$\sqrtsNN = 2.76\TeV$ \PbPb data collected during the 2011 LHC heavy ion run,
corresponding to an integrated luminosity of 2.3\mubinv, is also analyzed for
comparison purposes. The triggers,  event selection, and track reconstruction
are identical to those used in Ref.~\cite{Chatrchyan:2013nka}.

In order to select high-multiplicity \pPb collisions,
dedicated high-multiplicity triggers were implemented using the CMS level-1
and high-level trigger (HLT) systems.
The online track reconstruction at the HLT is based on
the three layers of pixel detectors, and requires a track origin within a cylindrical region of
length 30\unit{cm} along the beam axis and radius 0.2\unit{cm} perpendicular to the beam axis, centered at the nominal interaction point.
For each event, the vertex reconstructed with the highest number of pixel tracks is selected.
The number of pixel tracks (${N}_\text{trk}^\text{online}$)
with $\abs{\eta} < 2.4$, $\pt > 0.4 \GeVc$, and a distance of closest approach to this vertex of 0.4\unit{cm} or
less, is determined for each event. Several high-multiplicity ranges are defined with
prescale factors that are progressively reduced until, for the highest multiplicity events, no prescaling was applied.

In the offline analysis, hadronic collisions are selected by requiring a
coincidence of at least one HF tower containing more than 3\GeV of
total energy on either side of the interaction region.
Only towers within $3.0 < \abs{\eta} < 5.0$ are used in order to avoid the edges of the HF acceptance.
The \pPb interactions were simulated with both the \textsc{epos lhc}~\cite{Porteboeuf:2010um}
and the {\HIJING 1.383}~\cite{Gyulassy:1994ew} event generators. The requirement of having at least one primary
particle with total energy $E > 3.0$\GeV in each of the $\eta$ ranges $-5.0 < \eta < -3.0$ and $3.0 < \eta < 5.0$
is found to select 97--98\% of the total inelastic hadronic cross section.

Events in the offline analysis are also required to contain at least one reconstructed primary
vertex within 15\unit{cm} of the nominal interaction point along the beam axis ($z_{\text{vtx}}$)
and within 0.15\unit{cm} transverse to the beam trajectory.
At least two reconstructed tracks are required to be associated
with the primary vertex. Beam-related background is suppressed
by rejecting events for which less than 25\% of all reconstructed tracks
pass the track selection criteria for this analysis.
The \pPb instantaneous luminosity provided by the LHC in 2013
resulted in an approximately 3\% probability of at least one additional interaction
occurring in the same bunch crossing.
Such pileup events become more significant as the event multiplicity increases.
Following the procedure developed in Ref.~\cite{Chatrchyan:2013nka} for rejecting pileup events,
a 99.8\% purity of single-interaction events is achieved for the \pPb collisions
belonging to the highest multiplicity class of this analysis.

The CMS ``high-quality" tracks described in Ref.~\cite{Chatrchyan:2014fea}
are used in this analysis. Additionally, a reconstructed track is only considered as a
candidate track from the primary vertex if the significance of the separation along the beam axis ($z$)
between the track and the best vertex, $d_z/\sigma(d_z)$, and the significance
of the track impact parameter measured transverse to the beam,
$d_{\text{T}}/\sigma(d_{\text{T}})$, are each less than 3. The relative uncertainty
in \pt, $\sigma(\pt)/\pt$, is required to be less than 10\%.
To ensure high tracking efficiency and to reduce the rate of incorrectly reconstructed tracks,
only tracks within $\abs{\eta} < 2.4$ and with $\pt > 0.3 \GeVc$  are used in the analysis.
The entire \pPb data set is divided into classes of reconstructed track multiplicity, \noff,
where primary tracks with $\abs{\eta} < 2.4$ and $\pt > 0.4$\GeVc are counted.
A different \pt cutoff of 0.4\GeVc is used in the multiplicity determination because of the constraints
on the online processing time for the HLT.
The multiplicity classification in this analysis is identical to that used in Ref.~\cite{Chatrchyan:2013nka},
where more details are provided, including a table relating \noff to the fraction of minimum bias triggered events.

The peripheral \PbPb data collected during the
2011 LHC heavy ion run with a minimum bias trigger are also reanalyzed in order
to compare directly the \pPb and \PbPb systems in the same \noff ranges~\cite{Chatrchyan:2013nka}.
This \PbPb sample is reprocessed using the same event selection and track reconstruction
as for the present \pPb analysis.
A description of the 2011 \PbPb data set can be found in Ref.~\cite{Chatrchyan:2012xq}.  The correspondence between the \PbPb \noff values and the
total energy deposited in the HF~\cite{Chatrchyan:2012ta}, as characterized by
a collision centrality, is given in Ref.~\cite{Chatrchyan:2013nka}, ranging from 67\% centrality for \noff = 120 to 55\% centrality for \noff = 300.
\section{Analysis}

\subsection{Scalar product method}

In previous publications, CMS has analyzed the elliptic~\cite{Chatrchyan:2012ta} and higher-order~\cite{Chatrchyan:2013kba} flow coefficients for \PbPb collisions at $\sqrtsNN = 2.76\TeV$ using the ``traditional" event plane method~\cite{Poskanzer:1998yz}. It is now known that fluctuations in the participant geometry lead to $v_{n}$ coefficients that can vary event-by-event, with the average coefficients $\left\langle {{v_n}} \right\rangle $ being smaller than the corresponding root-mean-square values, $\sqrt {\left\langle {v_n^2} \right\rangle }$. The $v_{n}$ values found using the traditional event plane method will
fall somewhere between these two limits~\cite{Luzum:2012da}.  The scalar product method~\cite{Adler:2002pu,Luzum:2012da}, which is used in this paper, avoids this ambiguity and gives results that correspond to $\sqrt {\left\langle {v_n^2} \right\rangle }$~\cite{Luzum:2012da}.

The event plane angles can be expressed in terms of Q-vectors.  For a perfect detector response, the Q-vector corresponding to the $n$th-order azimuthal asymmetry for a given event is defined as

\begin{linenomath*}
\begin{equation}
\begin{split}
\label{eqn:qvector}
{\vec Q_n} = \left( {{Q_{nx}},{Q_{ny}}} \right) = \left( {\left| {{{\vec Q}_n}} \right|\cos \left( {n{\Psi _n}} \right),\left| {{{\vec Q}_n}} \right|\sin \left( {n{\Psi _n}} \right)} \right)  \\
 =\left( {\sum\limits_{i=1}^M {{w_i}\cos \left( {n{\phi _i}} \right),\sum\limits_{i=1}^M {{w_i}\sin \left( {n{\phi _i}} \right)} } } \right),
\end{split}
\end{equation}
\end{linenomath*}
where $M$ is the subevent multiplicity, $\phi_i$ is the azimuthal angle of the $i$th particle, $w_i$ are weighting factors,
and the corresponding event plane angle is given as
\begin{linenomath*}
\begin{equation}
\label{eqn:epqvec}
{\Psi _n} = {\frac{1}{n}}~{\tan ^{ - 1}}\left( {{\frac{{Q_{ny}}} {{Q_{nx}}}}} \right).
\end{equation}
\end{linenomath*}
Different weights $w_i$ are possible. For example, the Q-vectors with $w_i=1$  relate to the azimuthal particle density, with $w_i=p_{{\text{T}},i}$ to the
transverse momentum distribution, and with $w_i=E_{{\text{T}},i}$ to the transverse energy distribution.
Since the $v_n( \pt)$ coefficients increase
with \pt up to $\approx$3\GeVc, the choice of either \pt or $\ET$
weighting generally results in a better event plane angle resolution than a
unity particle weighting~\cite{Poskanzer:1998yz}.

Expressed in terms of complex weighted q-vectors, where
\begin{linenomath*}
\begin{equation}
{q_n} = {\frac{{\sum\limits_{i = 1}^M {{w_i}{\re^{in{\phi _i}}}} }} {W}},
\end{equation}
\end{linenomath*}
and $W = \sum\limits_{i = 1}^M {{w_i}} $,
the scalar product coefficients are found with
\begin{linenomath*}
\begin{equation}
\label{eqn:spa}
{v_n}\left\{ {{\text{SP}}} \right\} \equiv {\frac{\left\langle {{q_n}q_{n{\text{A}}}^*} \right\rangle }  {\sqrt {{\frac{\left\langle {q_{n{\text{A}}}^{}q_{n{\text{B}}}^*} \right\rangle \left\langle {q_{nA}^{}q_{n{\text{C}}}^*} \right\rangle } {\left\langle {q_{n{\text{B}}}^{}q_{n{\text{C}}}^*} \right\rangle }}} }}.
\end{equation}
\end{linenomath*}
In Eq.~(\ref{eqn:spa}), the weighted average $\langle\rangle$ for vectors $q_{n\alpha}$ and $q_{n\beta}$ with total weights $W_{\alpha}$ and $W_{\beta}$, where $\alpha$ and $\beta$ correspond to the second subscripts (if present) on the q-vectors in Eq.~(\ref{eqn:spa}),  is given by
\begin{linenomath*}
\begin{equation}
\label{eqn:spb}
\left\langle {q_{n\alpha}q_{n\beta}^*} \right\rangle  = \Re\left[ {\frac{\sum\limits_{i = 1}^{{N_{{\text{evt}}}}} {{W_{\alpha i}}{W_{\beta i}}{q_{n \alpha i}}q_{n \beta i}^*} }{\sum\limits_{i = 1}^{{N_{{\text{evt}}}}} {{W_{\alpha i}}{W_{\beta i}}}}} \right],
\end{equation}
\end{linenomath*}
\noindent where $N_{\text{evt}}$ is the total number of events. The A, B, and C subscripts in Eq.~(\ref{eqn:spa}), denoted using $\alpha$ and $\beta$ in Eq.~(\ref{eqn:spb}), refer to pseudorapidity ranges for which event planes are determined.
Here, the ``reference" event plane is the A plane, and the B and C planes are used to correct for the finite resolution of the A plane.
The q-vector with only one subscript, $q_n$ in Eq.~(\ref{eqn:spa}), is based on tracks within the specific \pt and $\eta$ range for which the azimuthal asymmetry coefficient is being measured. Unit weights are used in Eq.~(\ref{eqn:qvector}) in this case.

The two HF calorimeters are used to determine the A and B event planes, with the C plane established using the tracker.
In the HF detector regions, with $3.0 < \abs{\eta} < 5.0$, the sums in Eq.~(\ref{eqn:qvector}) are taken over the towers and the weights are taken as the transverse energy deposited in each tower, with no restriction placed on the tower energy.
For the tracker-based C plane, the sums are over the individual tracks with $0.3 < \pt < 3.0\GeVc$ and the weights are taken as the corresponding \pt values. The Q-vectors corresponding to event planes A, B, and C are ``recentered" to account for nonuniformities in the detector response~\cite{Poskanzer:1998yz, Barrette:1997pt}.  In recentering, the averages over all events of the x- and y-terms in Eq.~(\ref{eqn:qvector}) ($\left\langle {{Q_{nx}}} \right\rangle $
and
$\left\langle {{Q_{ny}}} \right\rangle $) are subtracted on an event-by-event basis when calculating
${\vec Q_n^{{\text{Recentered}}}}$.  That is,
\begin{linenomath*}
\begin{equation}
\vec Q_n^{{\text{Recentered}}} = \left( {{Q_{nx}} - \left\langle {{Q_{nx}}} \right\rangle ,\;{Q_{ny}} - \left\langle {{Q_{ny}}} \right\rangle } \right).
\end{equation}
\end{linenomath*}

The value of $q_n$ in Eq.~(\ref{eqn:spa}) is based on tracks within a specific \pt and $\eta$ range for which the azimuthal asymmetry coefficient is being measured. In this case, unit weights are used in Eq.~(\ref{eqn:qvector}) and no recentering corrections are applied.

It has been noted recently~\cite{Gardim:2012im,Bozek:2015bha,Heinz:2013bua,Xiao:2012uw}, and experimentally confirmed by CMS~\cite{Khachatryan:2015oea}, that the event plane angle should not be considered a global event observable.  In the CMS study~\cite{Khachatryan:2015oea}, the decorrelation between the event plane angles at pseudorapidity $\eta_{\text{A}}$ and $\eta_{\text{B}}$ is found to follow the functional form:
\begin{linenomath*}
\begin{equation}
\label{eqn:decor}
\cos \left[ {2\left\{ {{\Psi _n}\left( {{\eta_{\text{B}}}} \right) - {\Psi _n}\left( {{\eta_{\text{A}}}} \right)} \right\}} \right] = {\re^{ - F_n^\eta \left| {{\eta_{\text{B}}} - {\eta_{\text{A}}}} \right|}},
\end{equation}
\end{linenomath*}
\noindent where  $F_n^\eta$ is the decorrelation strength.

Such a decorrelation can arise from fluctuations of the geometry of the initial-state nucleons and their constituent partons~\cite{Gardim:2012im,Bozek:2015bha,Heinz:2013bua}. Previously it has been assumed that Fourier coefficients at pseudorapidity $\eta_{\text{ROI}}$, where ROI stands for ``region of interest", can be deduced using event plane angles found in a different pseudorapidity range (say, at $\eta_{\text{A}}$), with the caveat that a sufficient pseudorapidity gap is present to avoid short-range correlations.
The event plane angle found at $\eta_{\text{A}}$ is viewed as approximating a global participant plane angle set by the initial collision geometry and only differing from the ideal by its finite resolution, which, in turn, depends on both the number of particles used to define the angle and the azimuthal asymmetry  at $\eta_{\text{A}}$.
The event plane resolution is accounted for in Eq.~(\ref{eqn:spa})  by determining event planes in three separate regions of $\eta$ and assuming that these planes reflect the same underlying geometry, only differing by their respective resolutions.
The variation with pseudorapidity breaks this assumption and can have a significant effect on the harmonic coefficient values $v_{n}$ deduced using either the traditional or scalar product methods.

Considering event plane decorrelation, each of the scalar products in Eq.~(\ref{eqn:spa}) will be reduced by the decorrelation effect as indicated in Eq.~(\ref{eqn:decor}).
If the decorrelation strength $F_n$ remains relatively constant as a function of the pseudorapidity gap between event planes, the $v_n\{{\text{SP}}\}$ coefficient in the presence of decorrelation can be expressed in terms of the coefficient without decorrelation  $\bar v_n\{\text{SP}\}$ with
\begin{linenomath*}
\begin{equation}\begin{aligned}
{\upsilon _n}\left\{ {{\text{SP}}} \right\} &=
\frac{\left\langle {{q_n}q_{n{\text{A}}}^*} \right\rangle {\re^{ - F_n\left| {{\eta _{\text{A}}} - {\eta _{\text{ROI}}}} \right|}}}  {\sqrt {\frac{{\left\langle {{q_{n{\text{A}}}}q_{n{\text{B}}}^*} \right\rangle {\re^{ - F_n\left| {{\eta_{\text{A}}} - {\eta_{\text{B}}}} \right|}}\left\langle {{q_{n{\text{A}}}}q_{n{\text{C}}}^*} \right\rangle {\re^{ - F_n\left| {{\eta_{\text{A}}} - {\eta_{\text{C}}}} \right|}}}}{\left\langle {{q_{n{\text{B}}}}q_{n{\text{C}}}^*} \right\rangle {\re^{ - F_n\left| {{\eta_{\text{B}}} - {\eta_{\text{C}}}} \right|}}}} } \\
&= {\bar \upsilon_n}\left\{ {\text{SP}} \right\}\frac{{{\re^{ - F_n\left| {{\eta_{\text{A}}} - {\eta _{\text{ROI}}}} \right|}}} }{{{\re^{ - \frac{1}{2}F_n\left\{ {\left| {{\eta_{\text{A}}} - {\eta_{\text{B}}}} \right| + \left| {{\eta_{\text{A}}} - {\eta_{\text{C}}}} \right| - \left| {{\eta_{\text{C}}} - {\eta_{\text{B}}}} \right|} \right\}}}}} \\
&= {\bar \upsilon _n}\left\{ {\text{SP}} \right\}{\re^{ - F_n\left| {{\eta_{\text{C}}} - {\eta _{\text{ROI}}}} \right|}},
\end{aligned}\label{eqn:sp}
\end{equation}
\end{linenomath*}
\noindent
where $\eta_{\text{C}}$ is taken to fall between $\eta_{\text{A}}$ and $\eta_{\text{B}}$. Short-range, nonflow correlations, such as back-to-back dijets, resonance decay, etc., are again
suppressed by having a pseudorapidity gap between $\eta_{\text{ROI}}$ and $\eta_{\text{A}}$.

For the ``standard" analysis using a three subevent resolution correction where both the third subevent angle ($\Psi_{n}^{\text{C}}$) and the particles belonging to the region of interest are at midrapidity (${\eta _{\text{ROI}}} = {\eta_{\text{C}}} \approx 0$), it follows that the decorrelation effect will not strongly influence the deduced Fourier coefficient $v_{n}$.  It can be noted that the same result is expected if a two-subevent resolution correction is used, as is commonly done for symmetric collision systems.  However, if $\eta_{\text{ROI}}$ is different from $\eta_{\text{C}}$, the deduced $v_{n}$ value will be reduced by the decorrelation effect.

The pseudorapidity-dependent decorrelation of event planes can occur through different mechanisms.
Equation~(\ref{eqn:sp}) assumes a Gaussian decorrelation characterized by a fixed $F_{n}$ value.
It is also possible for $F_{n}^{\eta}$ to vary with $\eta$, in which case the $\eta$ dependence shown in Eq.~(\ref{eqn:decor}) and (\ref{eqn:sp}) would be more complicated.
A simplified MC simulation was used to explore the two Gaussian spreading scenarios, corresponding to a fixed or $\eta$-dependent $F_{n}^{\eta}$ factor.
It was found that the input $v_n$ values could be recovered by moving the $\Psi_n^{\text{C}}$ event plane along with the particles of interest.
An alternative source of decorrelation is the situation where rotation of the event plane angle results from a torque effect rather than a random spreading~\cite{Bozek:2015bha}.
In this case, the MC simulations showed that moving the $\Psi_n^{\text{C}}$ event plane does not fully correct for the decorrelation, although it does lead to results closer to the input values than is found by setting $\eta_{\text{C}} = 0$.
A comparison of the $v_2$ and $v_3$ results obtained with $\eta_{\text{C}}=0$ and with $\eta_{\text{C}} = \eta_{\text{ROI}}$ might help in estimating the relative importance of the different types of decorrelation possible in heavy ion collisions.
Event plane results using both of these assumptions for $\eta_{\text{C}}$ are reported.

Two different reference event planes are used in the analysis: HF$^{-}$ $(-5.0 < \eta < -3.0)$ and
HF$^{+}$ $(3.0 < \eta < 5.0)$. The corresponding resolution correction factors are determined with the
three subevent method where, for the HF$^{+}$(HF$^{-}$) reference plane (A-plane), the resolution correction is based on the HF$^{-}$(HF$^{+}$) event plane (B-plane) as well as either the midrapidity tracker event plane, with $-0.8 < \eta < 0.8$, or with event planes that correspond to the pseudorapidity range of the ROI (C-plane).
Since analyses where the midrapidity event plane $\eta_{\text{C}}$ is taken within $-0.8 < \eta_{\text{C}} < 0.8$ and analyses where $\eta_{\text{C}} = \eta_{\text{ROI}}$ are both presented,
the convention is adopted of  labelling results as ``$\eta_{\text{C}} = 0$" or ``$\eta_{\text{C}} = \eta_{\text{ROI}}$," respectively.

\subsection{Cumulant method}

If the particles emitted in a collision are correlated with a global reference frame, they will also be correlated with each other.
The cumulant method explores the collective nature of the anisotropic flow through the multiparticle correlations.
As the number of particles in the correlation study increases, the cumulant values will decrease if only part of the particle sample shares a common underlying symmetry, as would be the case for dijets.
The flow harmonics are studied using the Q-cumulant method~\cite{Bilandzic:2010jr}.
The $m$-particle ($m = 2$, 4, 6 or 8) $n$th-order correlators are first defined by
\begin{linenomath*}
\begin{equation}\begin{aligned}
	\dmean{2} &\equiv  \left<\!\left< \re^{in(\phi_{1} -
	  \phi_{2})} \right>\!\right>,\\
\dmean{4}&\equiv
	  \left<\!\left< \re^{in(\phi_{1} + \phi_{2} - \phi_{3}
	  - \phi_{4})} \right>\!\right>,\\
	\dmean{6}&\equiv
	  \left<\!\left< \re^{in(\phi_{1} + \phi_{2} + \phi_{3} - \phi_{4} - \phi_{5}
	  - \phi_{6})} \right>\!\right>,\\
\dmean{8}&\equiv
	  \left<\!\left< \re^{in(\phi_{1} + \phi_{2} + \phi_{3} + \phi_{4} - \phi_{5} - \phi_{6} - \phi_{7}
	  - \phi_{8})} \right>\!\right>,
\end{aligned}\label{eq:cumu}
\end{equation}
\end{linenomath*}
where $\phi_{i}$ is the azimuthal angle of the $i$th particle, and $\dmean{\dots}$ indicates that the average is taken over all $m$-particle combinations for all events.
In order to remove self-correlations, it is required that the $m$ particles be distinct.
The unbiased estimators of the reference $m$-particle cumulants~\cite{Bilandzic:2010jr}, $\cn{m}$, are defined as
\begin{linenomath*}
\begin{equation}
\begin{aligned}
	\cn{4} =& \dmean{4} - 2 \, \dmean{2}^2,\\
     \cn{6} =& \dmean{6} - 9 \, \dmean{4}\dmean{2} + 12 \, \dmean{2}^3,\\
	 \cn{8} =& \dmean{8} - 16 \, \dmean{6}\dmean{2} -18 \, \dmean{4}^2 \\
                  &+ 144 \, \dmean{4}\dmean{2}^2 - 144 \, \dmean{2}^4.
\end{aligned}\label{eq:cn}
\end{equation}
\end{linenomath*}
The reference flow $v_2\{m\}$ obtained by correlating the $m$ particles within the reference phase space of $\abs{\eta} < 2.4$ and \pt range of $0.3 < \pt < 3.0$\GeVc was presented in Ref.~\cite{Khachatryan:2015waa} using
\begin{linenomath*}
\begin{equation}
\begin{aligned}
	v_n\{4\} &= \sqrt[4]{-\cn{4}}\, \\
	v_n\{6\} &= \sqrt[6]{\cn{6} / 4}, \\
	v_n\{8\} &= \sqrt[8]{-\cn{8} / 33}.
\end{aligned}\label{eq:vncn}
\end{equation}
\end{linenomath*}
The cumulant calculations are done using the code described in Ref.~\cite{Bilandzic:2013kga}.

By replacing one of the particles in a correlator for each term in Eq.~(\ref{eq:cumu}) with a particle from certain ROI phase space in \pt or $\eta$, with the corresponding correlators denoted by primes, one can derive the differential
 $m$-particle cumulants as
\begin{linenomath*}
\begin{equation}\begin{aligned}
        d_n\{4\} =& \dmean{4^{\prime}} - 2 \dmean{2}\dmean{2^{\prime}},\\
         	d_n\{6\} =&\dmean{6^{\prime}} - 6 \dmean{2}\dmean{4^{\prime}} - 3 \dmean{2^{\prime}} \dmean{4} + 12 \dmean{2^{\prime}} \dmean{2}^2,\\
               d_n\{8\} =& \dmean{8^{\prime}} - 12 \dmean{2}\dmean{6^{\prime}} - 4 \dmean{2^{\prime}} \dmean{6} \\
                	& -18 \dmean{4^{\prime}} \dmean{4} + 72 \dmean{4} \dmean{2} \dmean{2^{\prime}} \\
                & + 72 \dmean{4^{\prime}} \dmean{2}^2 - 144 \dmean{2^{\prime}} \dmean{2}^3.
\end{aligned}\end{equation}
\end{linenomath*}
Then the differential $\vtwo{m}(\pt, \eta)$ can be extracted as
\begin{linenomath*}
\begin{equation}\begin{aligned}
	v_n\{4\}(\pt, \eta) &= -d_n\{4\} / (-c_n\{4\})^{3/4},\\
	\ 	v_n\{6\}(\pt, \eta) &= \frac{d_n\{6\}}{4} \Big/ \left(\frac{c_n\{6\}}{4}\right)^{5/6},\\
	\ 	v_n\{8\}(\pt, \eta) &= \frac{-d_n\{8\}}{33} \Big/ \left(\frac{-c_n\{8\}}{33}\right)^{7/8}.
\end{aligned}\end{equation}
\end{linenomath*}
An efficiency weight is applied to each track to account for detector nonuniformity and efficiency effects.
For this analysis,  the work of Ref.~\cite{Bilandzic:2013kga} was extended to allow for the explicit calculation of the differential Q-cumulants for the first time.
\subsection{Lee--Yang zeros method}

The LYZ method~\cite{Bhalerao:2003xf} allows for a direct study of the large-order behavior by using the asymptotic form of the
cumulant expansion to relate locations of the zeros of a generating function to the azimuthal correlations. This method has been employed in previous CMS \PbPb and \pPb analyses~\cite{Chatrchyan:2012ta,Chatrchyan:2013kba,Khachatryan:2015waa}. The $v_{2}$ harmonic averaged over $0.3 < \pt < 3.0\GeVc$ is found for each multiplicity bin using an integral generating function [17]. Similar to the cumulant methods, a weight for each track is implemented to account for detector-related effects.
Anisotropic flow is formally equivalent to a first-order phase transition. As a result, the first zero of the generating grand partition function can be viewed as anisotropic flow of the final-state system.

The integrated flow for the harmonic $n$ is the average value of the flow Q-vector projected onto the unit vector with angle $n{\Phi_{\text{R}}}$,
\begin{linenomath*}
\begin{equation}
{v_n^{\text{int}}} \equiv \left\langle {{Q_{nx}}\cos \left( {n{\Phi_{\text{R}}}} \right) + {Q_{ny}}\sin \left( {n{\Phi_{\text{R}}}} \right)} \right\rangle  = \left\langle {Q_n^{{\Phi_{\text{R}}}}} \right\rangle,
\end{equation}
\end{linenomath*}
where $\Phi_{\text{R}}$ is the actual reaction-plane angle. Since $\Phi_{\text{R}}$ is not an observable, the LYZ method is used to obtain an estimate of this
quantity.  In the present analysis, a complex product generating function is first defined  as
\begin{linenomath*}
\begin{equation}
  G_{n}^{\theta}(ir) = \left\langle {g_{n}^{\theta}(ir)}\right\rangle = \Bigl\langle{\prod_{j=1}^{M}[1+ ir\,w_{j}\cos{(n(\phi_{j}-\theta))}]}\Bigr\rangle,
\end{equation}
\end{linenomath*}
\noindent where $M$  is the event multiplicity, $\phi_{j}$ and $w_{j}$ are, respectively, the azimuthal angle and the weight of the $j$th particle, the average $\left\langle\right\rangle$ is taken over all events, and  $\theta$ is chosen to take discrete values within the range [0, $\pi/n)$ as
\begin{linenomath*}
\begin{equation}
  \theta = \frac{k}{n_{\theta}}\frac{\pi}{n}, \quad k= 0,1,2,...,n_{\theta}-1.
\end{equation}
\end{linenomath*}
\noindent The number of projection angles is set to $n_{\theta}$ = 5 to get the average values.
This number was found in the previous CMS studies to achieve convergence of the results~\cite{Chatrchyan:2012ta,Chatrchyan:2013kba,Khachatryan:2015waa}.

To calculate the yield-weighted integral flow, $G_{n}^{\theta}$ is evaluated for many values of the real positive variable $r$. Plotting the modulus
$|G_{n}^{\theta}(ir)|$ as a function of $r$, the integrated flow is directly related to the first minimum $r_{0}^\theta$ of the distribution, with
\begin{linenomath*}
\begin{equation}
v_n^{\theta,{\text{int}}} \left\{ \infty  \right\} \equiv {\frac{{j_{01}}} {r_0^\theta }},
\end{equation}
\end{linenomath*}
\noindent where ${j_{01}} \approx 2.405$ is the first root of the Bessel function $J_0(x)$.
The quoted results involve a final average over different $\theta$ values, with
\begin{linenomath*}
\begin{equation}
  v_{n}^{\text{int}}=\frac{1}{n_{\theta}}\sum_{\theta=0}^{n_{\theta}-1}v^{\theta, {\text{int}}}_{n} \left\{ \infty  \right\}.
\end{equation}
\end{linenomath*}

After the integrated flow coefficient $v_{n}^{\text{int}}$ is determined, the \pt- and $\eta$-dependent $v_{2}\{\text{LYZ}\}$ values are found using
\begin{linenomath*}
\begin{equation}
\frac{v_{n}^{\theta}}{v_{n}^{\theta, {\text{int}}}}=\Re \frac{\left\langle{g^{\theta}(ir^{\theta}_{0})\frac{cos(n(\phi_{j}-\theta))}{1+ir^{\theta}_{0}w_{j}\cos{(n(\phi_{j}-\theta))}}}\right\rangle_{\phi}}
{\left\langle{g^{\theta}(ir^{\theta}_{0})\sum_{j}\frac{w_{j}cos(n(\phi_{j}-\theta))}{1+ir^{\theta}_{0}w_{j}\cos{(n(\phi_{j}-\theta))}}}\right\rangle}.
\end{equation}
\end{linenomath*}
\noindent The average $\left\langle{...}\right\rangle_{\phi}$ in the numerator is taken  over the particles in the ROI. The average in the denominator is over all particles with $0.3 < \pt < 3.0 \GeVc$ and $\abs{\eta} < 2.4$.
Again, the final results involve an average over the different $\theta$ values
\begin{linenomath*}
\begin{equation}
  v_{n}=\frac{1}{n_{\theta}}\sum_{\theta=0}^{n_{\theta}-1}v^{\theta}_{n}.
\end{equation}
\end{linenomath*}

\subsection{Systematic uncertainties}

The systematic uncertainties resulting from the track selection and efficiency, from the vertex position, and from the pileup contamination contribute to all three methods (scalar product, cumulant, and LYZ).
The effects of track quality requirements were studied by varying the track selection requirements, $d_z/\sigma(d_z)$ and $d_{\text{T}}/\sigma(d_{\text{T}})$, from 2 to 5, and $\sigma(\pt)/\pt$ from 5\% to the case
where this requirement is not applied.
A comparison of the results using efficiency correction tables from \EPOS and \HIJING MC event generators was made to study the tracking efficiency uncertainty. By comparing the results from different event primary vertex positions along the beam direction, with $|z_{\text{vtx}}| < 3$\unit{cm} and $3 < |z_{\text{vtx}}| < 15$\unit{cm}, it is possible to investigate the uncertainties coming from the tracking acceptance effects.
The effects of pileup events were studied by looking at events where there was only one reconstructed vertex. The experimental systematic effects are found to have no significant dependence on \noff, \pt, or $\eta$.

The $v_2$ systematic uncertainties associated with the \PbPb collision results were found to be  comparable for the three methods (${\approx}3\%$), with  contributions from the track selection and efficiency (1--2\%), the
vertex position (1--2\%), and pileup effects (${<}1\%$).   Similar uncertainties are found for \pPb collisions
based on both the cumulant and scalar product methods.  For the LYZ \pPb results, a more conservative uncertainty of 11\% is quoted based on the large statistical uncertainties associated with the corresponding systematic studies.

In addition, a comparison was done between the results for the two different beam directions.
For the event plane analysis, the p-side and Pb-side HF detectors used to determine the event plane angles are switched by changing the beam direction.  Based on this study, where the small magnitude of the $v_3$ coefficient limits the statistical significance of the systematic studies, a larger, conservative systematic uncertainty is assigned to the $v_3\{\text{SP}\}$ results of $10\%$. The overall systematic uncertainties are summarized in Table~\ref{tab:syst}, and shown as grey boxes in the figures.

\begin{table}[ht]\renewcommand{\arraystretch}{1.2}\addtolength{\tabcolsep}{-1pt}
\centering
\topcaption{Systematic uncertainties.}
\label{tab:syst}
\begin{scotch}{ccccc }
&& $v_2(\pt)$ & $v_2(\eta)$ & $v_3$\\ \hline
\multirow{2}{*}{Scalar product}    & \pPb     & 3\% & 3\% & 10\%\\
& \PbPb    & 3\% & 3\% & 10\%\\[2ex]
\multirow{2}{*}{Cumulant}       & \pPb     & 3\% & 3\% & \NA\\
  & \PbPb    & 3\% & 3\% & \NA\\[2ex]
\multirow{2}{*}{Lee--Yang zeros} & \pPb     & 11\% & 11\% & \NA\\
 & \PbPb    & 3\% & 3\% & \NA \\
\end{scotch}
\end{table}

The multiparticle cumulant and LYZ analyses are expected to be relatively insensitive to nonflow effects.  For the
scalar product method, however, the nonflow effects can become significant as the differential particle density decreases, as is the situation
for the lower \noff ranges and for higher \pt values.   Also, the nonflow effects become
more significant as the gap between the primary
event plane ($\eta_{\text{A}}$) and the region of interest ($\eta_{\text{ROI}}$) becomes small.  In this paper, the nonflow influence on the scalar product results is viewed as part of the physics being explored and is not taken as a systematic uncertainty.

\section{Results}

We first explore the transverse momentum dependence of $v_2$ and $v_3$ in \pPb and \PbPb at comparable particle multiplicities.
The $v_2$ values were found using the scalar product, $m$-particle cumulant, and LYZ methods, denoted as $v_2\{\text{SP}\}$, $v_2\{m\}$, and $v_2\{\text{LYZ}\}$, respectively, while $v_3$ was found using only the scalar product method.

\begin{figure*}[hbt]
	\includegraphics[width=\linewidth]{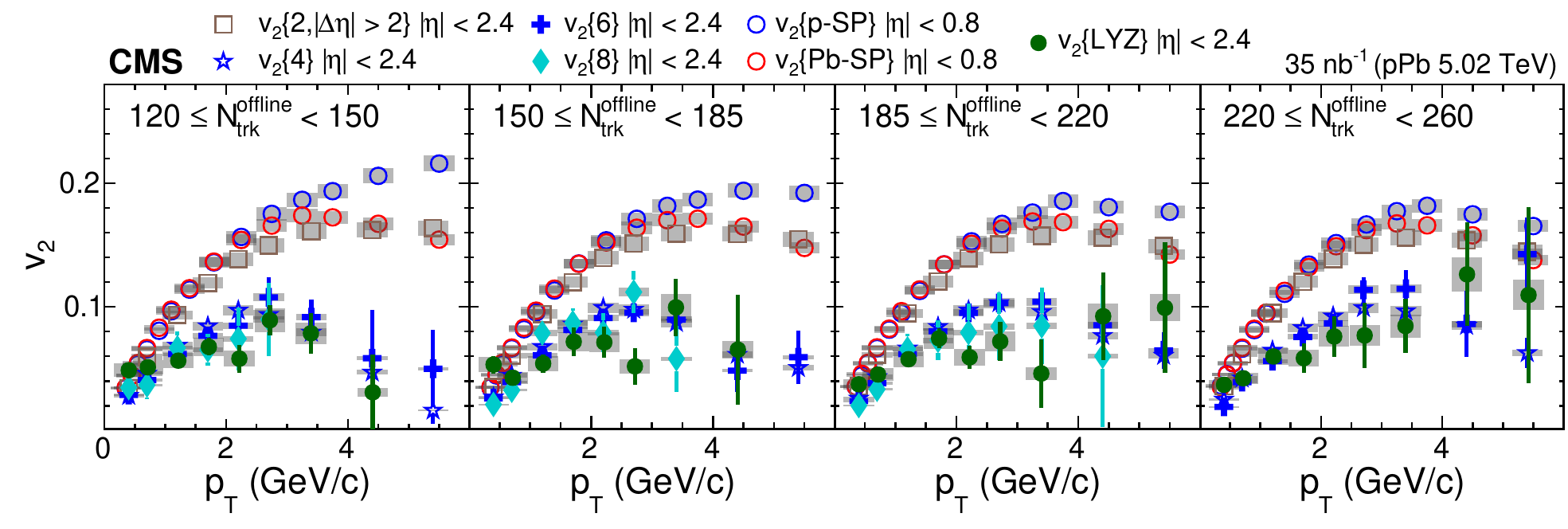}
	\includegraphics[width=\linewidth]{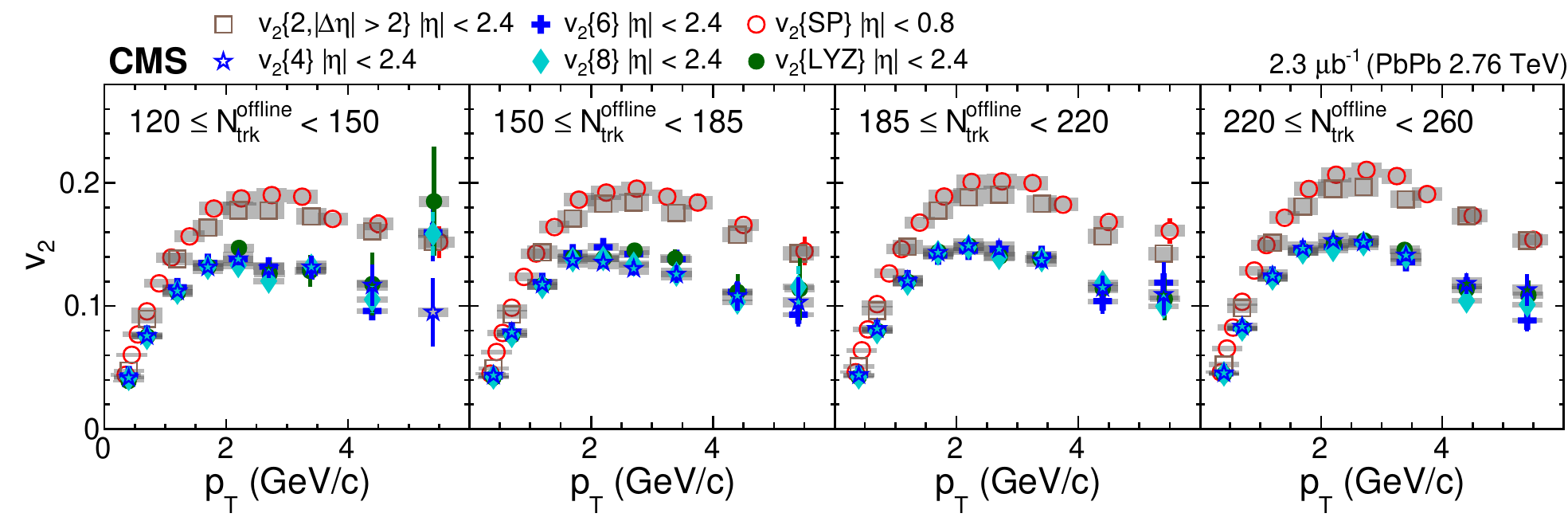}
	\caption{(Color online) (Top) The $v_2$ coefficients as a function of \pt in \pPb collisions for different \noff ranges. (Bottom) Same, but for \PbPb collisions.
		The $v_2\{2, \abs{\Delta\eta}>2\}$ and $v_2\{4\}$ results are from Ref.~\cite{Chatrchyan:2013nka}. For the \pPb collisions, the notations p-SP and Pb-SP indicate the pseudorapidity side of the reference event plane, and correspond to the p- and Pb-going directions, respectively. Pseudorapidities are given in the laboratory frame. Systematic uncertainties are indicated by the grey boxes. }
	\label{fig:v2cumu_pt}
\end{figure*}

The momentum-dependent $v_2(\pt)$ results in the region $\abs{\eta}<2.4$ for \pPb and \PbPb collisions are shown in Fig.~\ref{fig:v2cumu_pt}. The scalar product values, shown separately for the p- and Pb-going event planes, are found to be significantly higher than the multiparticle cumulant ($v_2\{4\}$, $v_2\{6\}$, and $v_2\{8\}$), and Lee--Yang zeros ($v_2\{{\text{LYZ}}\}$) results.
The two-particle correlations ($v_2\{2\}$) and lower-order cumulant ($v_2\{4\}$) measurements shown in the figure are from Ref.~\cite{Chatrchyan:2013nka}. As will be discussed when presenting the yield-weighted integral $v_2$ values, the greater values found for $v_2\{{\text{SP}}\}$ and $v_2\{2\}$ suggest a significant, and expected, contribution of fluctuations in the initial-state geometry to these results.

In the range of $\pt < 2\GeVc$ there is very little difference between the $v_2\{{\text{SP}}\}$ results obtained with the p- and Pb-going side event planes.   However, at higher transverse momenta, the p-going event plane leads to systematically larger values.  This behavior suggests that the nonflow contribution has a larger effect on the high-\pt $v_2$ values based on the p-going side event plane.  Monte Carlo simulations using the \HIJING event generator support a nonflow component to the $v_2$ signal that increases almost monotonically with \pt. In situations where both the event plane angle and the Q-vector associated with the region of interest are based on small numbers of particles, the nonflow behavior can be significant.    It is also possible that the \pt-dependent event-plane decorrelation effects might be different on the Pb- and p-going sides.

\begin{figure*}[thb]
\centering
\includegraphics[width=\linewidth]{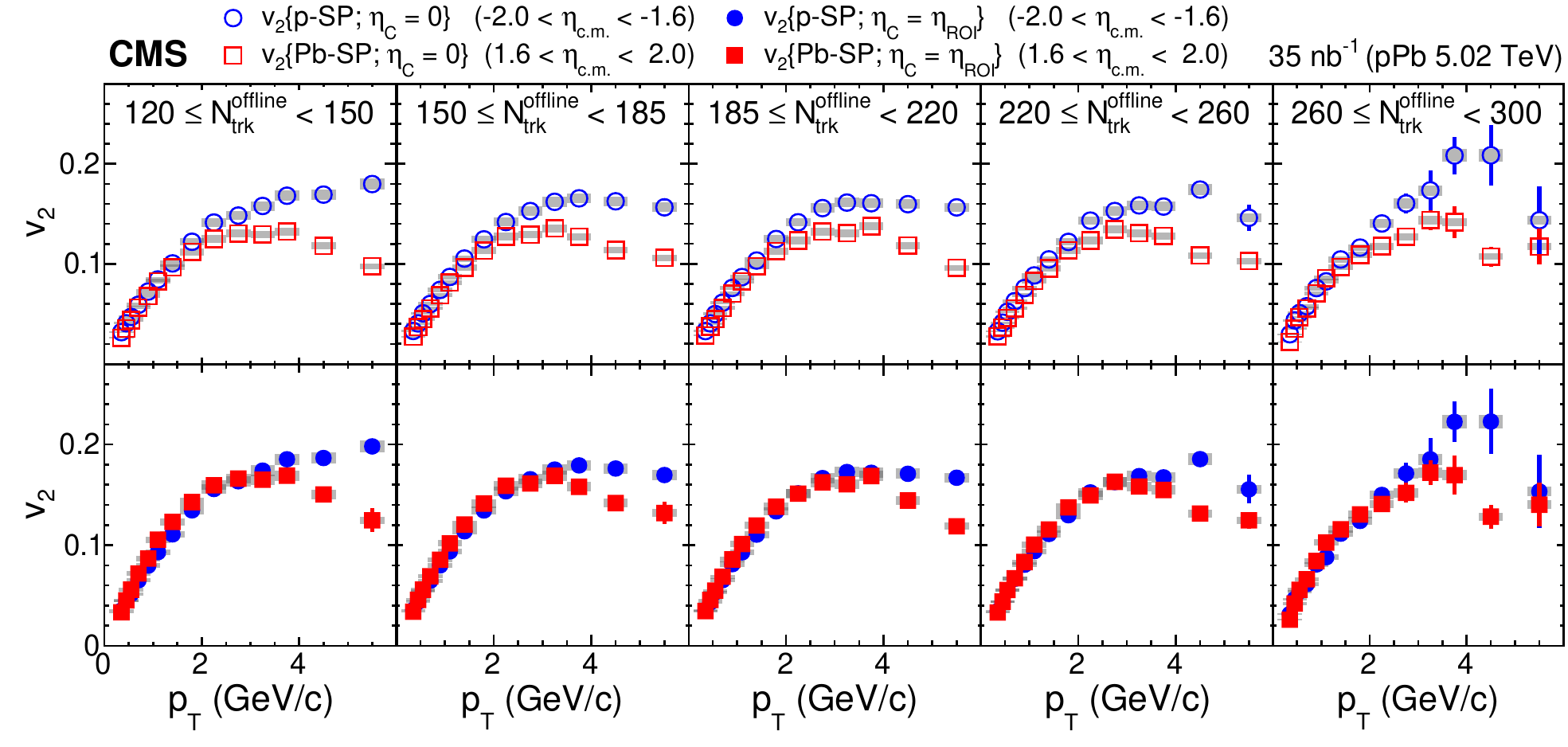}
\caption{(Color online) (Top) Comparison of $v_{2}(\pt)$ distributions located on the  Pb-going ($-2.0 < \eta_{\text{c.m.}} < -1.6$) and p-going ($1.6 < \eta_{\text{c.m.}} < 2.0$) sides of the tracker region, with $\eta_{\text{C}} = 0$.
The notations p-SP and Pb-SP indicate the pseudorapidity side of the reference event plane and correspond to the p- and Pb-going directions, respectively. (Bottom)  Same, but with $\eta_{\text{C}} = \eta_{\text{ROI}}$, as discussed in the text. Pseudorapidities are given in the laboratory frame. Systematic uncertainties are indicated by the grey boxes.}
\label{fig:EPpT_139_1}
\end{figure*}

In contrast to Fig.~\ref{fig:v2cumu_pt}, which uses an $\eta$ region that is symmetric in the lab frame, Fig.~\ref{fig:EPpT_139_1}  compares
the $v_2\{{\text{SP}}\}(\pt)$ results for symmetric pseudorapidity ranges in the center-of-mass frame.
The laboratory frame results for the range of $2.0 < \eta < 2.4$ correspond approximately to the center-of-mass range of $1.6 < \eta_{\text{c.m.}} < 2.0$ and are obtained with respect to the event plane found on the Pb-going side with $-5.0 < \eta < -3.0$, as indicated with the notation $v_2$~\{Pb-SP\}.  Similarly, the range of  $-1.6 < \eta < -1.2$ approximately corresponds to $-2.0 < \eta_{\text{c.m.}} < -1.6$. Here the results are obtained with respect to the event plane found on the p-going side with $3.0 < \eta < 5.0$, as indicated with the notation $v_2$~\{p-SP\}.  The measured values are shown separately with $\eta_{\text{C}} = 0$ and $= \eta_{\text{ROI}}$.
The reference event plane used in each case corresponds to the more distant HF detector.
In the region with $1.5 < \pt < 3.0\GeVc$, the enhancement observed on the Pb-going side ($-2.0 < \eta_{\text{c.m.}} < -1.6$; p-SP) with $\eta_{\text{C}} = 0$ (top row) is reduced by taking $\eta_{\text{C}} = \eta_{\text{ROI}}$ (bottom row).
This dependence on $\eta_{\text{C}}$ suggests the presence of event plane decorrelation.

\begin{figure*}[hbtp]
	\includegraphics[width=\linewidth]{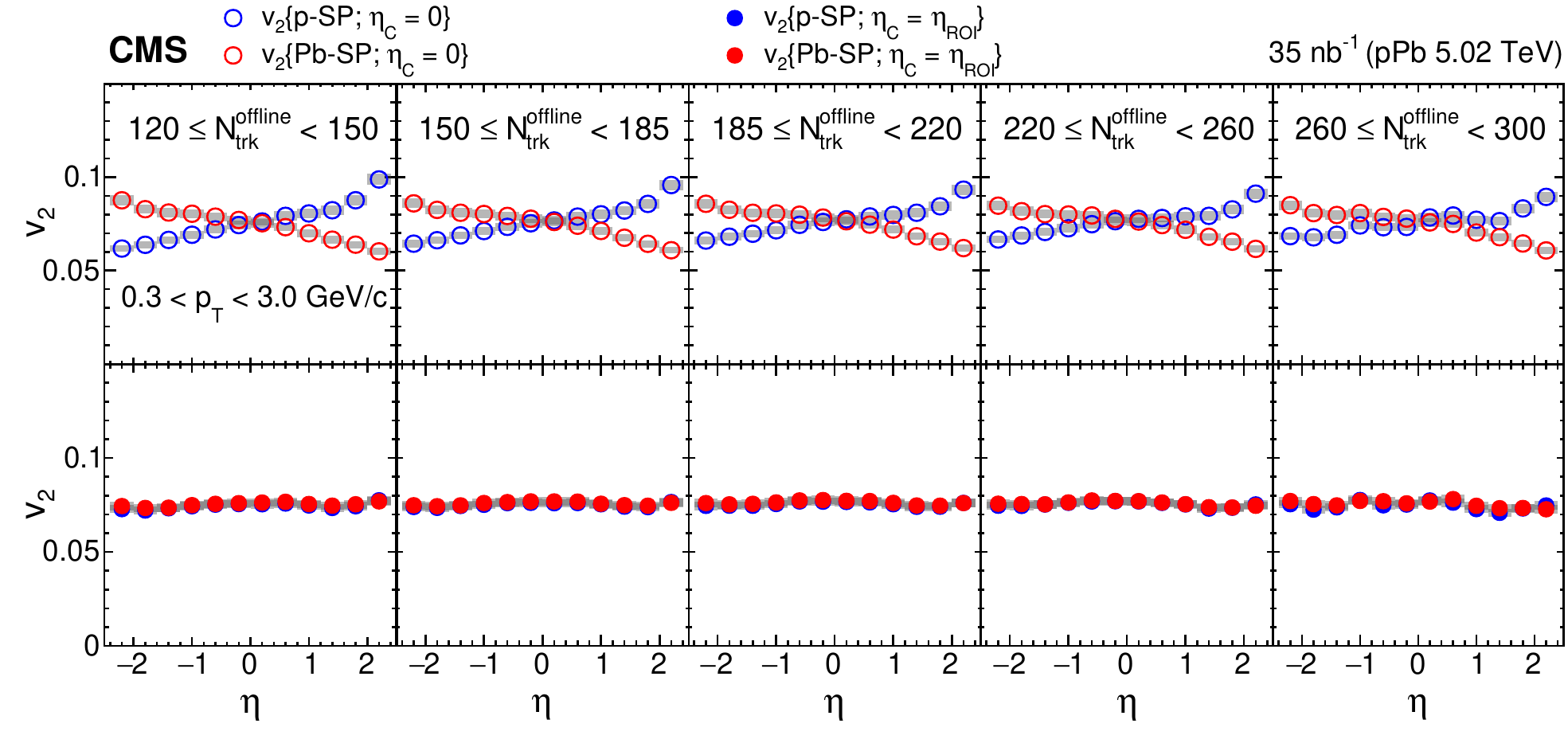}
	\caption {(Color online) (Top) Yield-weighted $v_2\{{\text{SP}}\}$ with $0.3 < \pt < 3.0\GeVc$ as a function of $\eta$ in \pPb  collisions for different \noff ranges with $\eta_{\text{C}} = 0$. (Bottom) Same, but with  $\eta_{\text{C}} = \eta_{\text{ROI}}$. The notations p-SP and Pb-SP indicate the pseudorapidity side of the reference event plane and correspond to the p- and Pb-going directions, respectively. Pseudorapidities are given in the laboratory frame. Systematic uncertainties are indicated by the grey boxes.}
	\label{fig:eta2all_139_1}
\end{figure*}

\begin{figure*}[hbtp]
	\includegraphics[width=\linewidth]{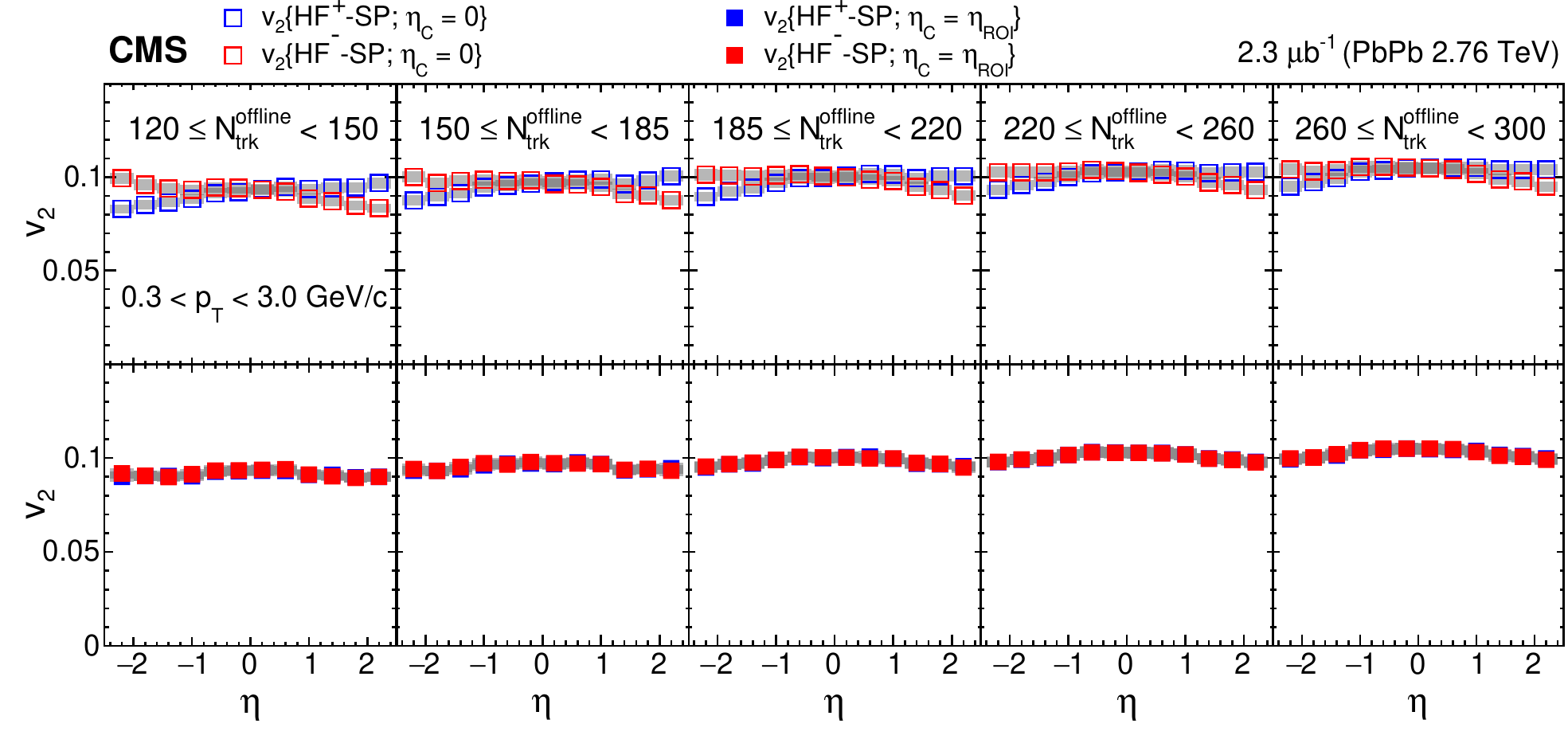}
	\caption{(Color online) (Top) Yield-weighted $v_2\{{\text{SP}}\}$ coefficients as a function of $\eta$ in  \PbPb collisions for different \noff ranges with $\eta_{\text{C}} =  0$. (Bottom) Same, but with
	$\eta_{\text{C}} = \eta_{\text{ROI}}$. The notations HF$^+$ and HF$^-$ indicate the pseudorapidity side of the reference event plane. Pseudorapidities are given in the laboratory frame. Systematic uncertainties are indicated by the grey boxes.}
	\label{fig:eta2all_109_1}
\end{figure*}

Further evidence for event plane decorrelation is seen by comparing the pseudorapidity dependence of the yield-weighted $v_2$ values for $0.3 < \pt < 3.0\GeVc$.  This is shown in Figs.~\ref{fig:eta2all_139_1} and \ref{fig:eta2all_109_1} for the \pPb and \PbPb collisions, respectively.  The top row in each figure shows the scalar product results with $\eta_{\text{C}} = 0$ and the bottom row with
 $\eta_{\text{C}} = \eta_{\text{ROI}}$. For the \pPb collisions, results are shown separately over the full pseudorapidity range of the CMS tracker using the HF event planes on the \Pp{}- and \Pb{}-going side of the collision.
For the symmetric \PbPb collisions, the results using the HF$^+$ and HF$^-$ event planes are shown separately.
The yield-weighted elliptic flow coefficients for \PbPb collision are found to be ${\approx}20\%$ larger than for \pPb collisions.
In the absence of decorrelation effects, the choice of $\eta_{\text{C}} = 0$ or  $= \eta_{\text{ROI}}$ would be expected to result in similar distributions. In previous \PbPb studies~\cite{Chatrchyan:2012ta,Chatrchyan:2013kba}, taking $\eta_{\text{C}} = 0$, the $v_2(\eta)$ values with $\eta<0$ were reported using the event plane with $3.0 <  \eta < 5.0$, and the values with $\eta > 0$ were reported using the event plane with $-5.0 < \eta < -3.0$, thus achieving the largest possible gap in pseudorapidity.
Before accounting for an increasing decorrelation of event planes with an increasing pseudorapidity gap, the $v_2$ values based on \Pp{}-going and \Pb{}-going side event planes (\pPb collisions) or HF$^{+}$ and HF$^{-}$ event planes (\PbPb collisions) show different pseudorapidity dependences, with the values decreasing as the gap with the reference event plane increases.  This reference event plane dependence largely disappears once a correction is applied for decorrelation effects, with the corrected $v_2$ values showing very little pseudorapidity dependence.  The resulting boost invariance is consistent with the azimuthal dependence being determined by the initial-state geometry.  For the \pPb collisions, the results  with $2.0 < \eta < 2.4$ determined using the p-going side reference event plane are systematically higher in each of the \noff ranges.  This is consistent with the reduced multiplicity associated with this eta region, allowing for an increased influence of  nonflow effects.

The current results suggest that event plane decorrelation effects might be significant in trying to understand the pseudorapidity dependence of the flow coefficients.  The results  with $2.0 < \eta < 2.4$ determined using the p-going side reference event plane are systematically higher, suggesting the possible influence of nonflow effects.

\begin{figure*}[hbtp]
	\includegraphics[width=\linewidth]{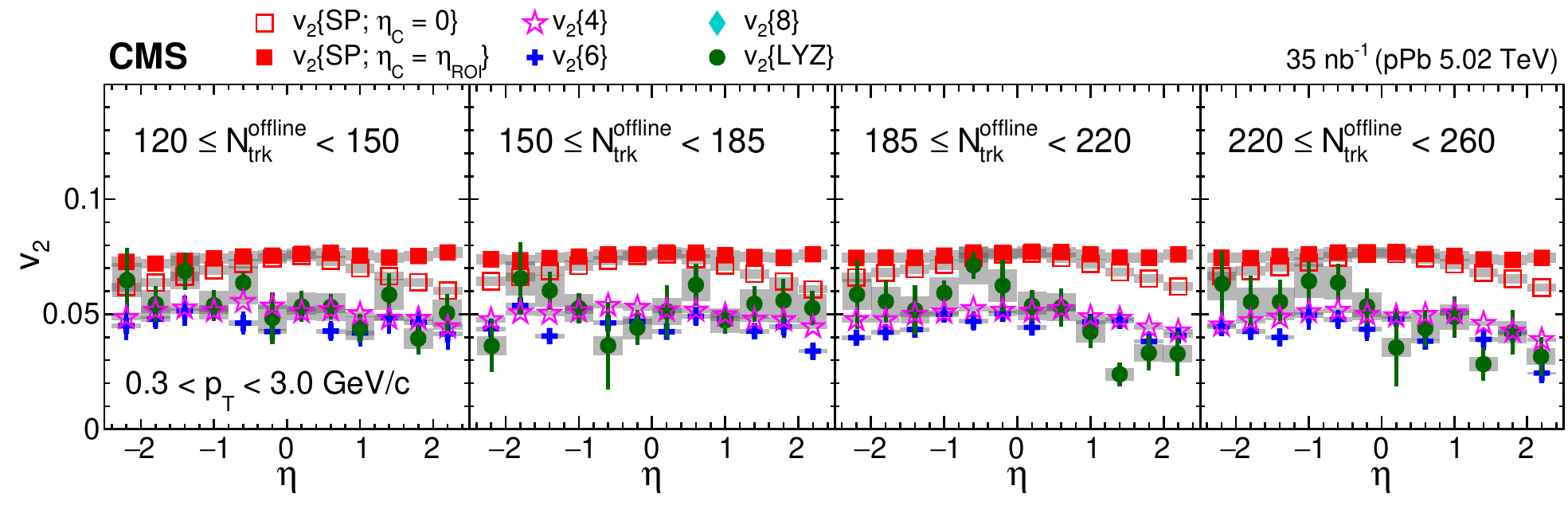}
	\includegraphics[width=\linewidth]{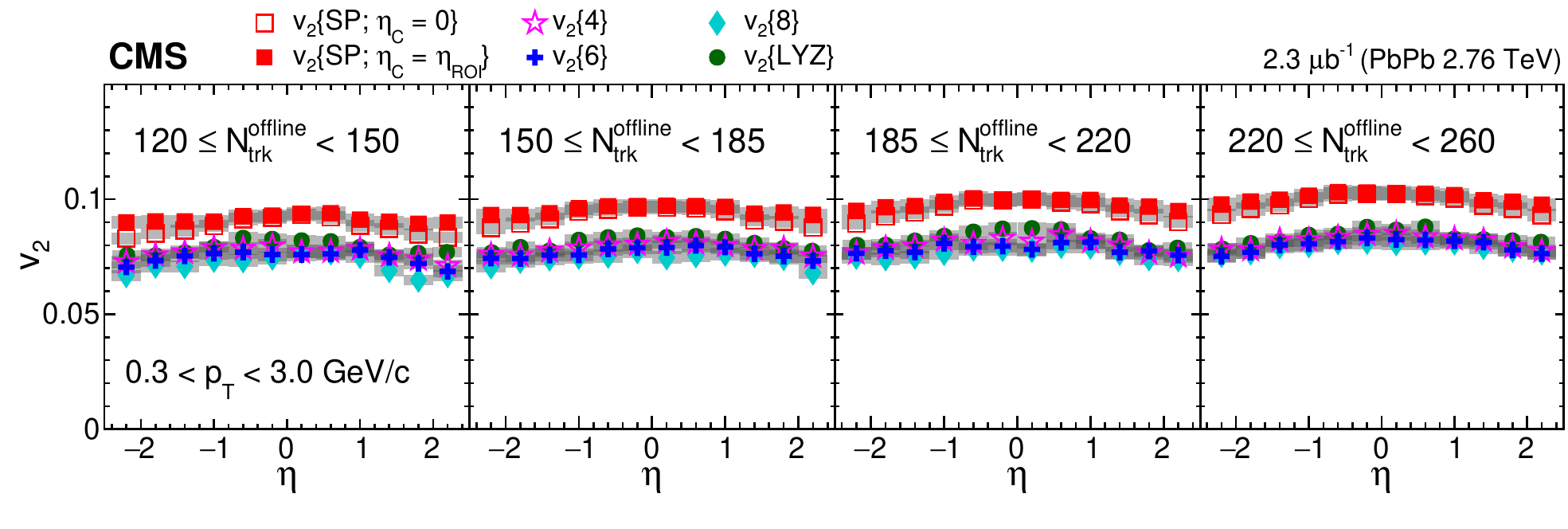}
	\caption{(Color online) (Top) Yield-weighted $v_2$ values calculated using the scalar product, cumulant, and LYZ methods as a function of $\eta$ in \pPb collisions for different \noff ranges. (Bottom) Same, but for \PbPb collisions. The $v_2\{{\text{SP}}\}$ results are based on the furthest HF event plane in pseudorapidity from the particles of interest. Pseudorapidities are given in the laboratory frame. Systematic uncertainties are indicated by the grey boxes.}
	\label{fig:v2cumu_eta}
\end{figure*}

Expanding on the results in Figs.~\ref{fig:eta2all_139_1} and~\ref{fig:eta2all_109_1}, which show only $v_2$ from the scalar product method, the yield-weighted average $v_2$ values for all of the analysis methods are shown in Fig.~\ref{fig:v2cumu_eta}.
It is interesting to note that the pseudorapidity dependence is almost flat for the scalar product calculations where  $\eta_{\text{C}} = \eta_{\text{ROI}}$.
This is in contrast to the scalar product results for $\eta_{\text{C}} = 0$ and for the higher-order particle correlation analyses, where the $v_2$ values at larger pseudorapidities are significantly smaller.
It is only for the scalar product analysis with $\eta_{\text{C}} = \eta_{\text{ROI}}$ that a partial accounting for the event plane decorrelation behavior is achieved.
Both the cumulant and LYZ analyses employ integral reference flows based on the full range of the CMS tracker and thus are not able to account for decorrelation effects.
There is an apparent asymmetry as a function of pseudorapidity for the LYZ results for the two highest \noff ranges, with a larger $v_2$ signal observed on the Pb-going side event plane.
Although this asymmetry appears to be larger than that found for the cumulant or scalar product analyses, the large statistical uncertainties make a direct comparison difficult.

It can be seen from Fig.~\ref{fig:v2cumu_eta} that the \PbPb results for a given \noff range are consistently higher than the corresponding \pPb results.  This likely reflects the very different collision geometries for the two systems, with the elliptic flow  for \PbPb collisions being influenced by the lenticular-shaped overlap region developed in non-central collisions of two Pb nuclei.  In a later discussion, this result will be contrasted with a similar comparison for the $v_3$ harmonic.

As already suggested for the \pt-dependent results, the difference between the scalar product and two-particle correlations results, as compared to the higher-order correlation studies, is likely to reflect initial-state fluctuation effects. Event-by-event fluctuations in the location of the participant nucleons can have a large and
method-dependent influence on the harmonic coefficients~\cite{Alver:2008zza,Ollitrault:2009ie}.
Expressing the fluctuations in terms of the azimuthal anisotropy in the participant plane $v$, where the harmonic number
is suppressed, the magnitude of the fluctuations is given by
$\sigma _v^2 \equiv \left\langle {{v^2}} \right\rangle  - {\left\langle v \right\rangle ^2}$.
To leading order in $\sigma_v$~\cite{Ollitrault:2009ie},
two- and four-particle correlations are affected differently, with
\begin{linenomath*}
\begin{equation}
\label{eqn:Eqn11}
v{\left\{ 2 \right\}^2} = \left\langle {{v^2}} \right\rangle  = {\left\langle v \right\rangle ^2} + \sigma _v^2
\end{equation}
\end{linenomath*}
and
\begin{linenomath*}
\begin{equation}
\label{eqn:Eqn12}
v{\left\{ 4 \right\}^2} = {\left( {2{{\left\langle {{v^2}} \right\rangle }^2} - \left\langle {{v^4}} \right\rangle } \right)^{1/2}} \approx {\left\langle v \right\rangle ^2} - \sigma _v^2.
\end{equation}
\end{linenomath*}
Multiparticle correlations with more than four particles are
expected to give results similar to those of four-particle correlations. Fluctuations affect the scalar product and two-particle correlations in a similar manner. The difference between the scalar product and higher-order cumulant results therefore reflects the initial-state fluctuations.

\begin{figure*}[thbp]
\centering
\includegraphics[width=\linewidth]{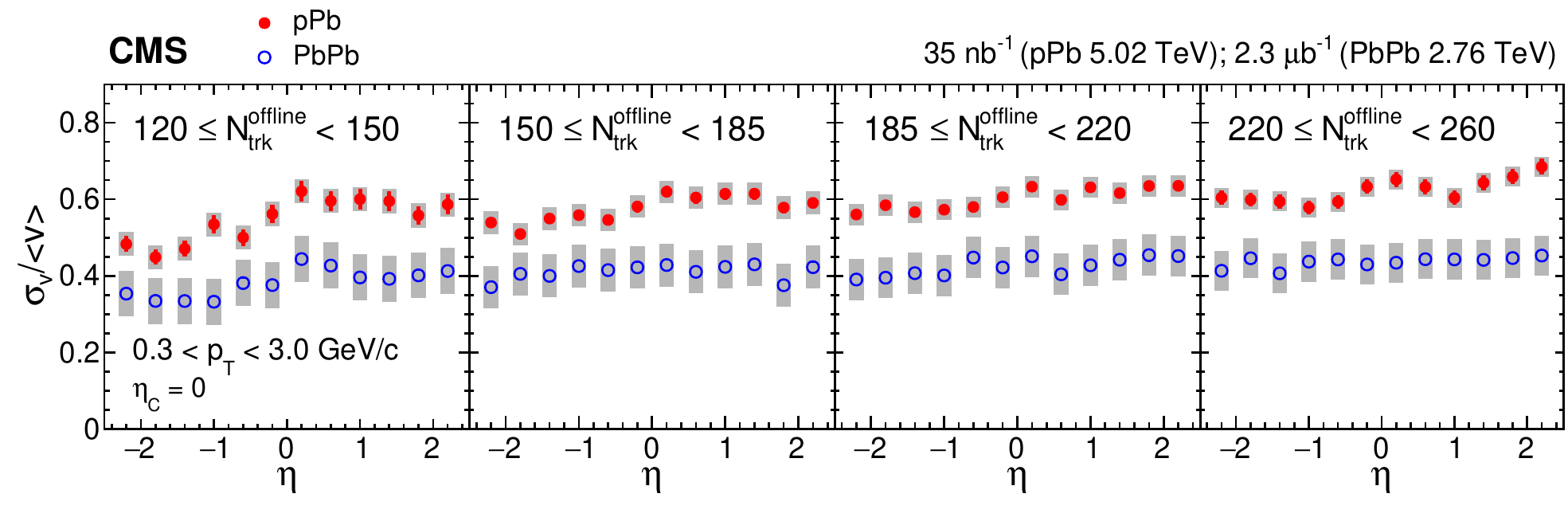}
\caption{(Color online) The ratio $\sigma_v$/$\langle v \rangle$ in the \pPb and \PbPb systems as a function of pseudorapidity for the indicated \noff ranges. Pseudorapidities are given in the laboratory frame. Systematic uncertainties are indicated by the grey boxes.}
\label{fig:Fluct}
\end{figure*}

Using Eqs.~(\ref{eqn:Eqn11}) and (\ref{eqn:Eqn12}), the fluctuation ratio $\sigma_v/\langle v\rangle$ can be calculated as
\begin{linenomath*}
\begin{equation}
\frac{{\sigma _v} }{ {\langle v\rangle}} = \sqrt {\frac{{{v_2}{{\left\{ 2 \right\}}^2} - {v_2}{{\left\{ 4 \right\}}^2}}}{{{v_2}{{\left\{ 2 \right\}}^2} + {v_2}{{\left\{ 4 \right\}}^2}}}} = \sqrt {\frac{{{v_2}{{\left\{ {\text{SP}} \right\}}^2} - {v_2}{{\left\{ 4 \right\}}^2}} }{ {{v_2}{{\left\{ {\text{SP}} \right\}}^2} + {v_2}{{\left\{ 4 \right\}}^2}}}} .
\end{equation}
\end{linenomath*}
This ratio is shown in Fig.~\ref{fig:Fluct} for the \pPb and \PbPb collisions in different \noff ranges.  The $v_2\{{\text{SP}}\}$ results with $\eta_{\text{C}} = 0$ are used in the calculations since the $v_2\{4\}$ results are expected to be affected by decorrelation effects. The fluctuation component is found to be significantly larger for the \pPb collisions as compared to the \PbPb results. A small (15--20\%) increase in the ratio is found for both the \pPb and \PbPb systems as the \noff range increases.  The \pPb system also shows an increase in the ratio as the
pseudorapidity increases.

\begin{figure}[thbp]
\centering
\includegraphics[width=0.5\textwidth]{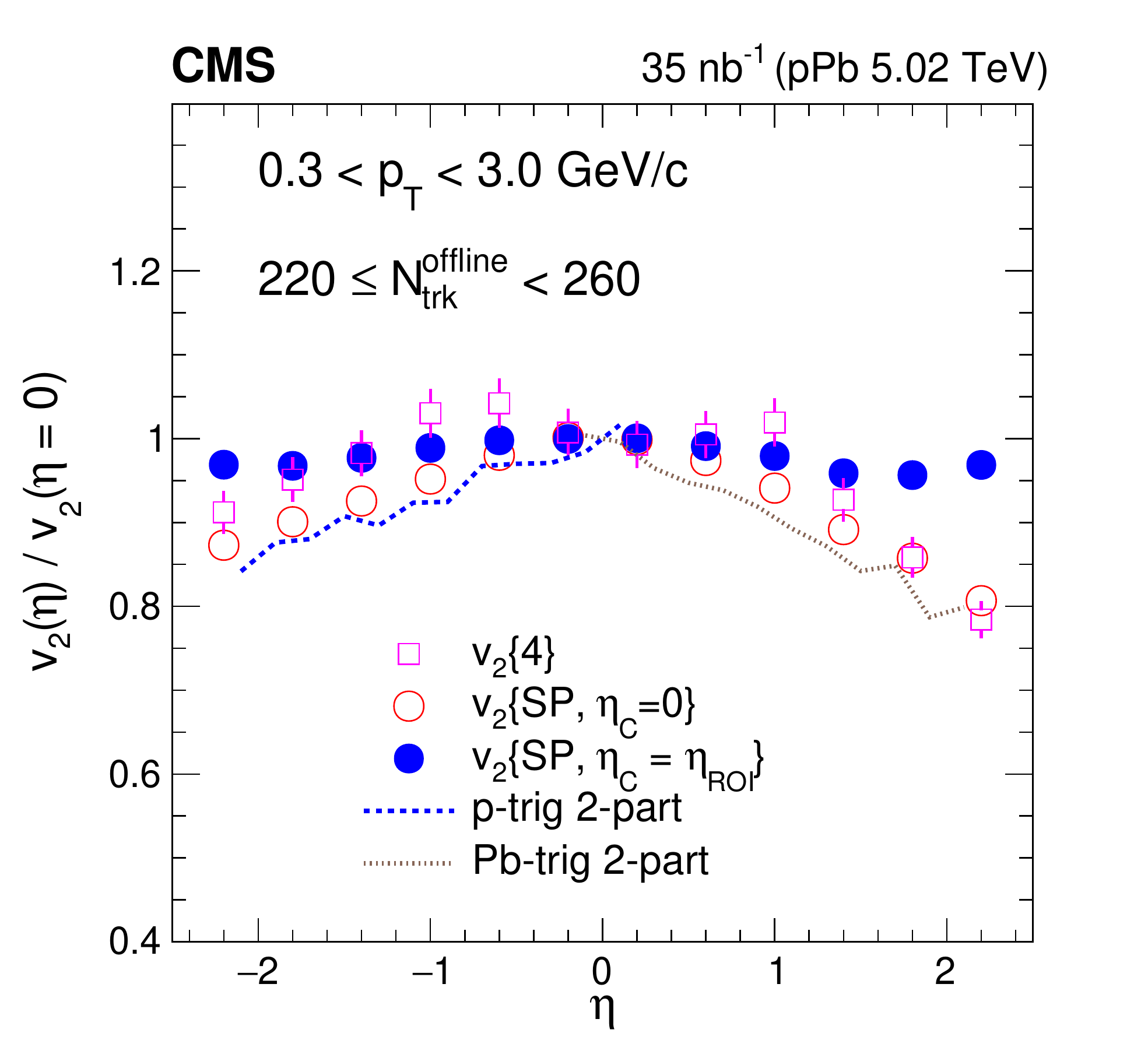}
\caption{(Color online) Comparison of the scalar product ($v_2\{{\text{SP}}\}$) and cumulant ($v_2\{4\}$) results for the ratio $v_{2}(\eta)/v_{2}(\eta = 0)$ with the two-particle correlation results from Ref.~\cite{Khachatryan:2016ibd} for \pPb collisions at $\sqrtsNN = 5.02\TeV$ and with $220\leq \noff < 260$.  The scalar product results with $\eta < 0$ use the p-side reference event plane with $3.0 < \eta < 5.0$, and the results with $\eta > 0$ are based on the Pb-side reference event plane with $-5.0 < \eta < -3.0$. The two-particle correlation results of Ref.~\cite{Khachatryan:2016ibd} for p-side (p-trig 2-part) and Pb-side (Pb-trig 2-part) trigger particles are shown without the peripheral $v_{2}$ component subtraction, a correction for nonflow effects that increases the $v_{2}$ harmonics. Pseudorapidities are given in the laboratory frame.
Error bars are statistical uncertainties.}
\label{fig:norm_pPb}
\end{figure}

The results presented here can be used to evaluate in more detail previous CMS analyses
which suggest a significant pseudorapidity dependence of the $v_{2}$ coefficient of \pPb collisions, with a larger ``flow" signal on the Pb-going side~\cite{Khachatryan:2016ibd}.
That study was based on a two-particle correlation analysis and focused on the ratio $v_{2}(\eta)/v_{2}(\eta = 0)$.
Since the Ref.~\cite{Khachatryan:2016ibd} analysis does not take into account decorrelation effects, it is most closely related to the  scalar product analysis with $\eta_{\text{C}} = 0$ and to the multiparticle correlation measurements based on the integral flow coefficients found using an extended range of the CMS tracker acceptance.
The Ref.~\cite{Khachatryan:2016ibd} results are compared to the scalar product and four-particle cumulant results in Fig.~\ref{fig:norm_pPb}.
Agreement is found among these measurements.
The scalar product results with $\eta_{\text{C}} = \eta_{\text{ROI}}$, also shown in Fig.~\ref{fig:norm_pPb}, fall off more slowly when moving away from midrapidity.

\begin{figure*}[hbtp]
	\includegraphics[width=\linewidth]{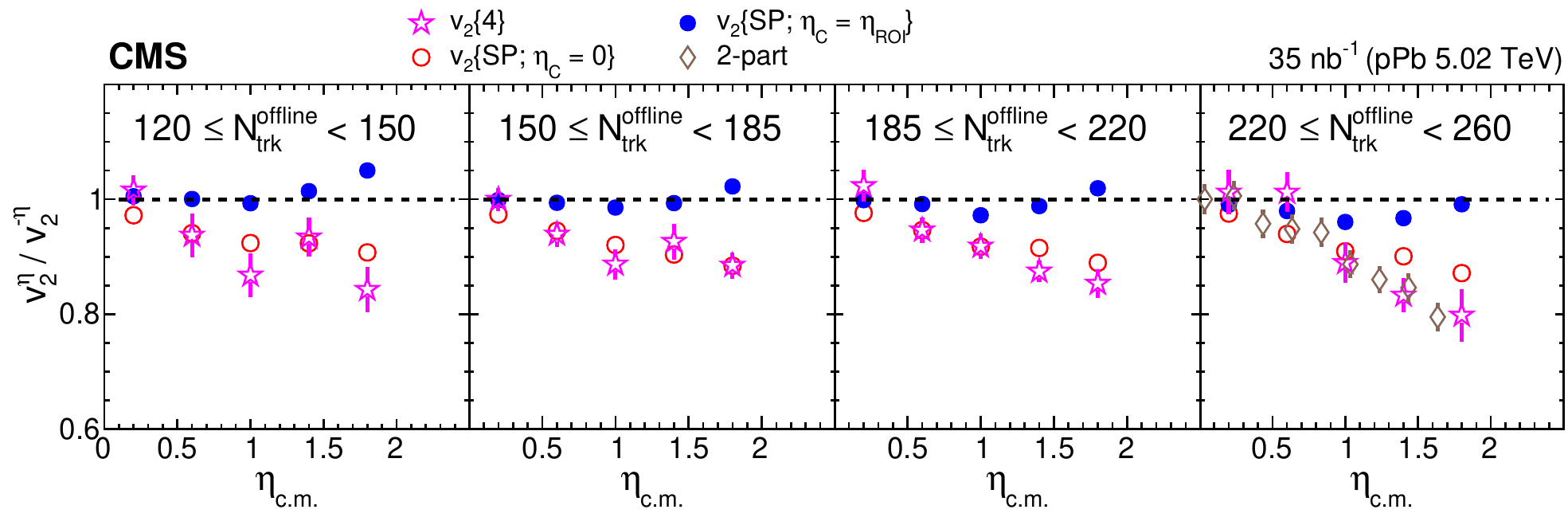}
	\caption{(Color online) Ratio of the p- to Pb-going side $v_2$ coefficients at comparable  $\eta_{\text{c.m.}}$  values for \pPb collisions. The two-particle correlation results (labelled ``2-part") are from
	Ref.~\cite{Khachatryan:2016ibd}. The reference $HF$ event plane is the one furthest from the particles of interest.}
	\label{fig:ratio}
\end{figure*}

\begin{figure*}[thbp]
	\includegraphics[width=\linewidth]{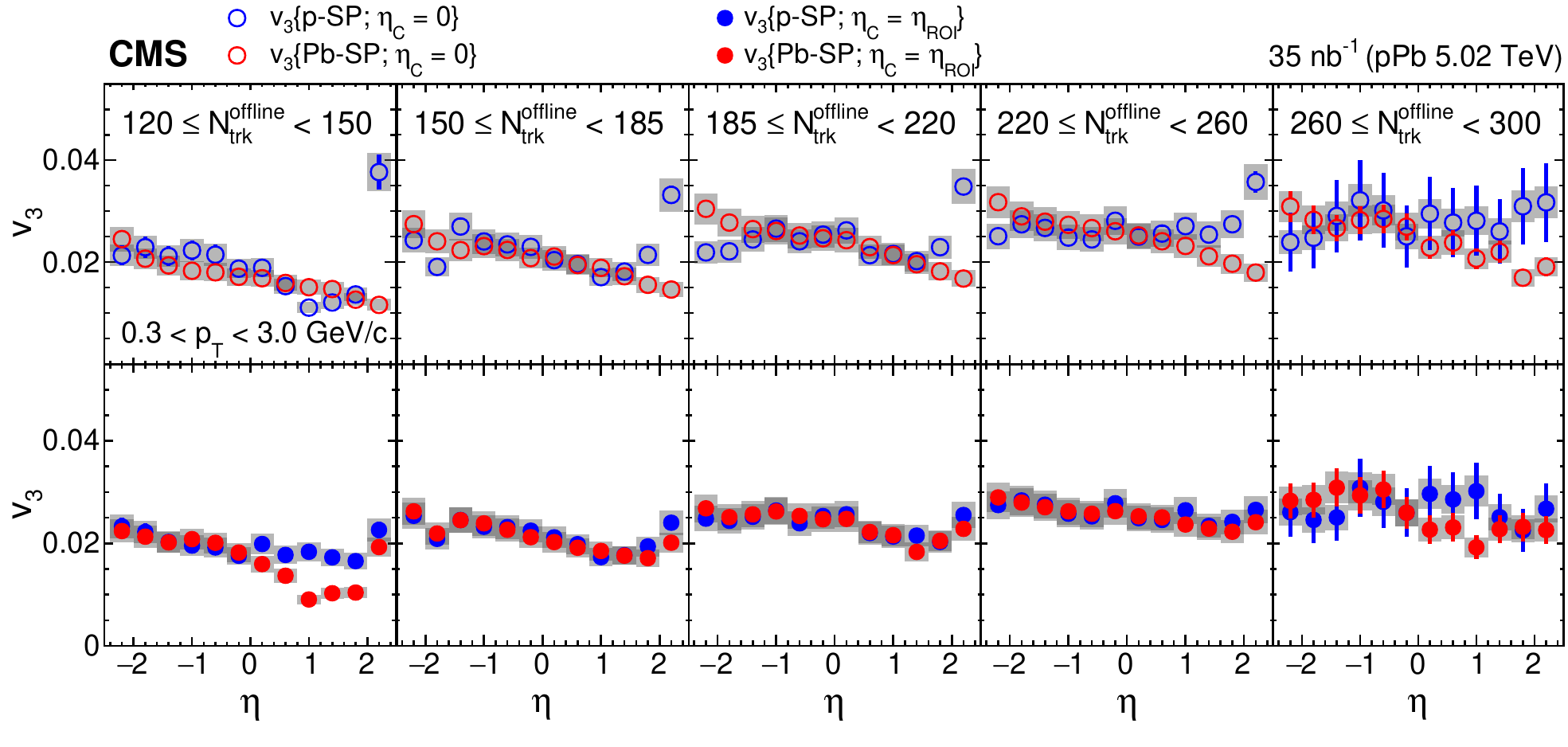}
\caption{(Color online) (Top) The $v_{3}$ values from the scalar product method for \pPb collisions at  $\sqrtsNN = 5.02\TeV$ with $\eta_{\text{C}} = 0$. (Bottom) Same, but with $\eta_{\text{C}} = \eta_{\text{ROI}}$. The notations p-SP and Pb-SP indicate the pseudorapidity side of the reference event plane and correspond to the p- and Pb-going directions, respectively. Pseudorapidities are given in the laboratory frame. Systematic uncertainties are indicated by the grey boxes.}
\label{fig:eta3all_139_1}
\end{figure*}

To explore further the possible asymmetry in the pseudorapidity-dependent $v_2$ results of Fig.~\ref{fig:v2cumu_eta} for the \pPb system, Fig.~\ref{fig:ratio} shows the ratios of the yield-weighted integral values on the p- and Pb-going sides at comparable center-of-mass pseudorapidity for \pPb collisions.
The results are shown for the scalar product analyses with $\eta_{\text{C}} = 0$ and $= \eta_{\text{ROI}}$,  and for the four-particle cumulant analysis.
Also shown are the comparable results from the Ref.~\cite{Khachatryan:2016ibd} analysis.
For the \pPb results where decorrelation effects are not taken into account (\ie, $v_2\{{\text{SP}},\eta_{\text{C}} = 0\}$ and $v_2\{4\}$), the Pb-going side values are significantly larger.
The asymmetry between the Pb-going and p-going sides largely disappears when decorrelation effects are taken into account.
A small asymmetry continues to be present when decorrelation effects are considered (\ie,  $v_2\{{\text{SP}},\eta_{\text{C}} = \eta_{\text{ROI}}\}$), although it needs to be recognized that the procedure of moving the $\eta_{\text{C}}$ range with $\eta_{\text{ROI}}$ is not expected to fully account for these effects if a torque-effect decorrelation is present; there may be some additional influence of nonflow effects
when the $\eta$ gap between the $\eta_{\text{C}}$  and either the $\eta_{\text{A}}$ or $\eta_{\text{B}}$ event planes becomes small.

In contrast to the second order Fourier coefficients discussed above, triangular flow, corresponding to
the $v_3$ Fourier harmonic, is believed to arise from fluctuations in the participant geometry in collisions of heavy nuclei.
It is interesting to see how this behavior extends to the very asymmetric \pPb system.
Fig.~\ref{fig:eta3all_139_1} shows the scalar product results for the \pPb collisions at $\sqrtsNN = 5.02\TeV$ with $\eta_{\text{C}} = 0$ (top) and $= \eta_{\text{ROI}}$ (bottom), respectively, as a function of $\eta$. Yield-weighted $v_3$ values with $0.3 < \pt < 3.0\GeVc$ are shown.
A pronounced jump in $v_3$, which becomes smaller with increasing \noff, is observed for $\eta > 2$ when using the p-going side reference event plane.
This could be due to nonflow effects when the ROI is close to the reference event plane.
For the Pb-going side reference event plane, a similar, but much smaller effect, may be present when taking $\eta_{\text{C}} = \eta_{\text{ROI}}$.

\begin{figure*}[thbp]
	\includegraphics[width=\linewidth]{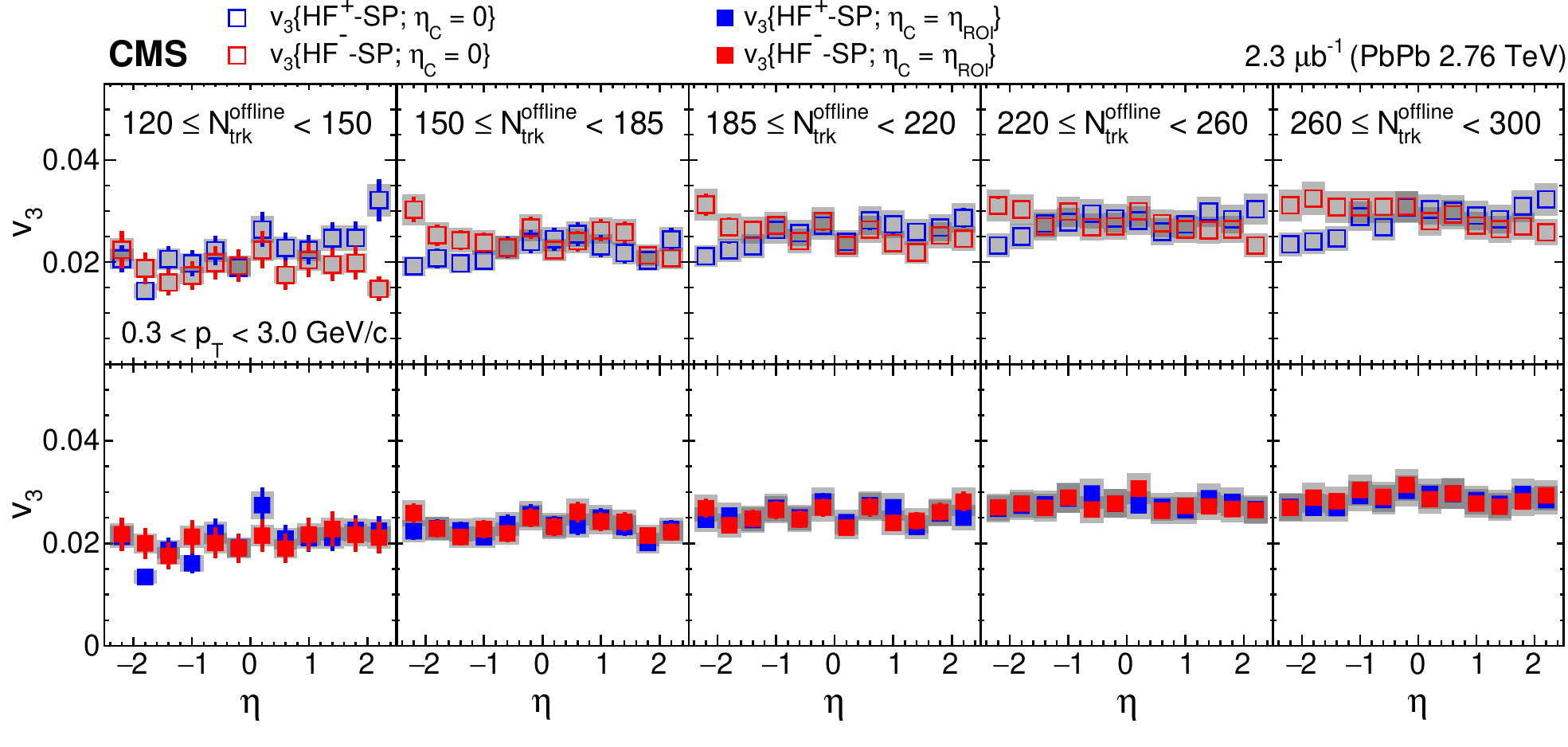}
\caption{(Color online) (Top) The $v_{3}$ values from the scalar product method for \PbPb collisions at  $\sqrtsNN = 2.76\TeV$ with $\eta_{\text{C}} = 0$. (Bottom) Same, but with $\eta_{\text{C}} = \eta_{\text{ROI}}$.  The notations HF$^+$ and HF$^-$ indicate the pseudorapidity side of the reference event plane. Pseudorapidities are given in the laboratory frame. Systematic uncertainties are indicated by the grey boxes.}
\label{fig:eta3all_109_1}
\end{figure*}

A small pseudorapidity dependence is
seen in the $v_3\{\eta_{\text{C}} = \eta_{\text{ROI}}\}$ results, with the values becoming smaller on the p-going side. This might suggest a changing level of fluctuations driving the triangular flow signal.  The pseudorapidity dependence appears to become less significant as \noff increases.
Fig.~\ref{fig:eta3all_109_1} shows the corresponding scalar product results for the \PbPb collisions at $\sqrtsNN = 2.76\TeV$ with $\eta_{\text{C}} = 0$ (top) and $= \eta_{\text{ROI}}$ (bottom).
The $v_3$ values are found to increase with increasing \noff for both systems, as previously observed in Ref.~\cite{Chatrchyan:2013nka}.  However, contrary to what is found for
the $v_2$ coefficients,  the $v_3$ values are very similar for the \pPb and \PbPb systems in a given \noff range.

\begin{figure*}[thbp]
	\includegraphics[width=\linewidth]{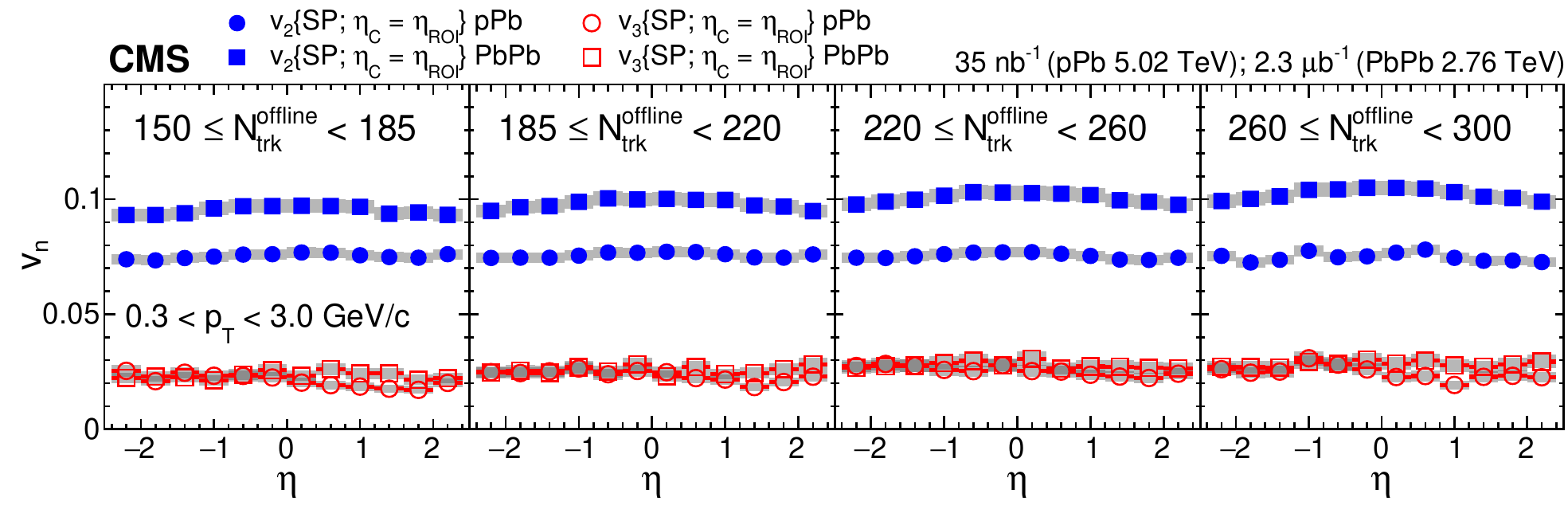}
	\caption{(Color online) The $v_{2}$ and $v_{3}$ values for \pPb (\PbPb) collisions at $\sqrtsNN = 5.02 (2.76)\TeV$ with $\eta_{\text{C}} = \eta_{\text{ROI}}$. The $v_n\{{\text{SP}}\}$ results are based on the furthest HF event plane in pseudorapidity.  Pseudorapidities are given in the laboratory frame. Systematic uncertainties are indicated by the grey boxes.}
\label{fig:eta3}
\end{figure*}

In order to show the system dependence of $v_2$ and $v_3$ more directly, Fig.~\ref{fig:eta3} shows scalar product results
with $\eta_{\text{C}} = \eta_{\text{ROI}}$ for both the \pPb and \PbPb systems.
The $v_3$ values, believed to result almost entirely from initial geometry fluctuations, are almost the same for the two systems.
The $v_2$ values are still likely to reflect the lenticular shape of the collision geometry in the \PbPb system, leading to larger $v_2$ coefficients than seen for the \pPb system. The \PbPb $v_2$ values are also found to increase with increasing event activity, reflecting the additional contribution of the changing collision overlap geometry.

\section{Summary}

The pseudorapidity and transverse momentum dependencies of the elliptic flow $v_{2}$ coefficient are presented for \pPb collisions at $\sqrtsNN= 5.02\TeV$
and for peripheral \PbPb collisions at $\sqrtsNN= 2.76\TeV$ based on scalar product,  multiparticle cumulant, and Lee--Yang zeros analyses.
The data are obtained using the CMS detector. The $\eta$ dependence of the triangular flow $v_{3}$ coefficient is also presented based on the scalar product analysis.
For the first time, \pt- and $\eta$-dependent cumulant results are presented based on 6- and 8-particle correlations.
The results provide detailed information for the theoretical understanding of the initial state effect and final state evolution mechanism.

All methods lead to a similar $\eta$ dependence for the $v_2$ harmonic across the pseudorapidity range studied.
The scalar product results are consistently higher than the corresponding multiparticle correlation behavior, with the $v_{2}\{4\}$,  $v_{2}\{6\}$, $v_{2}\{8\}$, and $v_{2}\{\text{LYZ}\}$ having comparable magnitude.
An analysis of fluctuations suggests their greater influence in the system formed in \pPb as compared to that in the \PbPb collisions.
No significant pseudorapidity dependence is found for the fluctuation component, although there is a small increase in the level of the fluctuations with increasing \noff in both the \pPb and \PbPb systems.  The boost invariance indicated by the decorrelation-corrected results confirms that the flow signal develops very early in the collision and thus reflects the initial-state geometry.

A method is presented to account for the possible decorrelation of the event plane angle with an increasing $\eta$ gap between two regions of pseudorapidity. The results suggest that most of the $\eta$ dependence observed using the different methods might be a consequence of the decorrelation effect.  Earlier results exploring the $\eta$ dependence of elliptic flow in heavy ion collisions may need to be reassessed based on the presence of such decorrelation effects.

Only a small difference is found for the $v_{2}$ coefficients on the Pb- and p-going sides for the \pPb collisions once decorrelation effects are considered.
This is in contrast to a previous study, in which the decorrelation effects were not considered and where a larger $v_{2}$ value was found on the Pb-going side.  If the decorrelation effects are not considered, as is the case with the current cumulant, LYZ, and scalar product analysis with $\eta_{\text{C}} = 0$, good agreement is found with the previous results.
When decorrelation effects are considered, there appears to be very little longitudinal dependence of the flow coefficients near midrapidity.

The yield-weighted $v_2$ results of \pPb and \PbPb collisions at comparable values of \noff show a similar $\eta$ dependence,
with the heavier system values being about 20\% higher than found for \pPb collisions.
No significant difference is observed for the \PbPb $v_3$ values as compared to \pPb collisions, suggesting that the $v_3$ results are solely a consequence of fluctuations in the initial-state participant geometry.

\begin{acknowledgments}
We congratulate our colleagues in the CERN accelerator departments for the excellent performance of the LHC and thank the technical and administrative staffs at CERN and at other CMS institutes for their contributions to the success of the CMS effort. In addition, we gratefully acknowledge the computing centers and personnel of the Worldwide LHC Computing Grid for delivering so effectively the computing infrastructure essential to our analyses. Finally, we acknowledge the enduring support for the construction and operation of the LHC and the CMS detector provided by the following funding agencies: BMWFW and FWF (Austria); FNRS and FWO (Belgium); CNPq, CAPES, FAPERJ, and FAPESP (Brazil); MES (Bulgaria); CERN; CAS, MoST, and NSFC (China); COLCIENCIAS (Colombia); MSES and CSF (Croatia); RPF (Cyprus); SENESCYT (Ecuador); MoER, ERC IUT, and ERDF (Estonia); Academy of Finland, MEC, and HIP (Finland); CEA and CNRS/IN2P3 (France); BMBF, DFG, and HGF (Germany); GSRT (Greece); OTKA and NIH (Hungary); DAE and DST (India); IPM (Iran); SFI (Ireland); INFN (Italy); MSIP and NRF (Republic of Korea); LAS (Lithuania); MOE and UM (Malaysia); BUAP, CINVESTAV, CONACYT, LNS, SEP, and UASLP-FAI (Mexico); MBIE (New Zealand); PAEC (Pakistan); MSHE and NSC (Poland); FCT (Portugal); JINR (Dubna); MON, RosAtom, RAS, RFBR and RAEP (Russia); MESTD (Serbia); SEIDI, CPAN, PCTI and FEDER (Spain); Swiss Funding Agencies (Switzerland); MST (Taipei); ThEPCenter, IPST, STAR, and NSTDA (Thailand); TUBITAK and TAEK (Turkey); NASU and SFFR (Ukraine); STFC (United Kingdom); DOE and NSF (USA).

\hyphenation{Rachada-pisek} Individuals have received support from the Marie-Curie program and the European Research Council and Horizon 2020 Grant, contract No. 675440 (European Union); the Leventis Foundation; the A. P. Sloan Foundation; the Alexander von Humboldt Foundation; the Belgian Federal Science Policy Office; the Fonds pour la Formation \`a la Recherche dans l'Industrie et dans l'Agriculture (FRIA-Belgium); the Agentschap voor Innovatie door Wetenschap en Technologie (IWT-Belgium); the Ministry of Education, Youth and Sports (MEYS) of the Czech Republic; the Council of Science and Industrial Research, India; the HOMING PLUS program of the Foundation for Polish Science, cofinanced from European Union, Regional Development Fund, the Mobility Plus program of the Ministry of Science and Higher Education, the National Science Center (Poland), contracts Harmonia 2014/14/M/ST2/00428, Opus 2014/13/B/ST2/02543, 2014/15/B/ST2/03998, and 2015/19/B/ST2/02861, Sonata-bis 2012/07/E/ST2/01406; the National Priorities Research Program by Qatar National Research Fund; the Programa Severo Ochoa del Principado de Asturias; the Thalis and Aristeia programs cofinanced by EU-ESF and the Greek NSRF; the Rachadapisek Sompot Fund for Postdoctoral Fellowship, Chulalongkorn University and the Chulalongkorn Academic into Its 2nd Century Project Advancement Project (Thailand); the Welch Foundation, contract C-1845; and the Weston Havens Foundation (USA).

\end{acknowledgments}

\bibliography{auto_generated}

\cleardoublepage \appendix\section{The CMS Collaboration \label{app:collab}}\begin{sloppypar}\hyphenpenalty=5000\widowpenalty=500\clubpenalty=5000\vskip\cmsinstskip
\textbf{Yerevan~Physics~Institute,~Yerevan,~Armenia}\\*[0pt]
A.M.~Sirunyan, A.~Tumasyan
\vskip\cmsinstskip
\textbf{Institut~f\"{u}r~Hochenergiephysik,~Wien,~Austria}\\*[0pt]
W.~Adam, F.~Ambrogi, E.~Asilar, T.~Bergauer, J.~Brandstetter, E.~Brondolin, M.~Dragicevic, J.~Er\"{o}, M.~Flechl, M.~Friedl, R.~Fr\"{u}hwirth\cmsAuthorMark{1}, V.M.~Ghete, J.~Grossmann, N.~H\"{o}rmann, J.~Hrubec, M.~Jeitler\cmsAuthorMark{1}, A.~K\"{o}nig, I.~Kr\"{a}tschmer, D.~Liko, T.~Madlener, I.~Mikulec, E.~Pree, D.~Rabady, N.~Rad, H.~Rohringer, J.~Schieck\cmsAuthorMark{1}, R.~Sch\"{o}fbeck, M.~Spanring, D.~Spitzbart, J.~Strauss, W.~Waltenberger, J.~Wittmann, C.-E.~Wulz\cmsAuthorMark{1}, M.~Zarucki
% ...
\vskip\cmsinstskip
% ...
1:~Also at~Vienna~University~of~Technology,~Vienna,~Austria\\
% ...
\end{sloppypar}
\end{document}